\documentclass[manuscript,nonacm]{acmart}

\usepackage{algorithmic}
\usepackage{subcaption}
\usepackage{multirow}
\usepackage{enumitem}
\usepackage{csquotes}
\usepackage{amsmath} 
\usepackage{pifont}
\usepackage{tabularx}
\usepackage[most]{tcolorbox}

\newcommand{\hc}{\paragraph{\textcolor{red}{Human cooperation}}}
\newcommand{\pa}{\paragraph{\textcolor{blue}{Partial automation}}}

\newcommand{\ParticipantOne}{\textbf{P1$_{P}$}}
\newcommand{\ParticipantTwo}{\textbf{P2$_{P}$}}

\newcommand{\ParticipantFour}{\textbf{P3$_{P}$}}
\newcommand{\ParticipantFive}{\textbf{P4$_{C}$}}
\newcommand{\ParticipantSix}{\textbf{P5}$_{CE}$}
\newcommand{\ParticipantSeven}{\textbf{P6}$_{E}$}
\newcommand{\ParticipantEight}{\textbf{P7}$_{PC}$}
\newcommand{\ParticipantNine}{\textbf{P8}$_{PC}$}
\newcommand{\ParticipantTen}{\textbf{P9}$_{PC}$}
\newcommand{\ParticipantEleven}{\textbf{P10}$_{PC}$}
\newcommand{\ParticipantTwelve}{\textbf{P11}$_{CE}$}
\newcommand{\ParticipantThirteen}{\textbf{P12}$_{CE}$}
\newcommand{\ParticipantFourteen}{\textbf{P13}$_{E}$}
\newcommand{\ParticipantFifteen}{\textbf{P14}$_{E}$}

\definecolor{green}{RGB}{5,150,105}
\definecolor{red}{RGB}{220,38,38}
\definecolor{orange}{RGB}{234,88,12}

\definecolor{lik1}{rgb}{0.36, 0.06, 0}
\definecolor{lik2}{rgb}{0.6, 0.0, 0}
\definecolor{lik3}{rgb}{0.9, 0.57, 0.22}
\definecolor{lik4}{rgb}{1, 0.85, 0.4}
\definecolor{lik5}{rgb}{0.58, 0.77, 0.49}

\definecolor{theme1}{HTML}{F78C6B}
\definecolor{theme2}{HTML}{FFD166}
\definecolor{theme3}{HTML}{06D6A0}
\definecolor{theme4}{HTML}{118AB2}
\definecolor{theme5}{HTML}{073B4C}
\definecolor{theme6}{HTML}{EF476F}

\definecolor{mygreen}{rgb}{0, 0.5, 0}
\definecolor{myred}{rgb}{0.85, 0.375, 0.055}

\newcommand{\quots}[1]{``#1"}

\begin{document}

\title[Shared Control for Game Accessibility]{Shared Control for Game Accessibility: Understanding Current Human Cooperation Practices to Inform the Design of Partial Automation Solutions}

\author{Dragan Ahmetovic}
\email{dragan.ahmetovic@unimi.it}
\orcid{0000-0001-5745-1230}
\affiliation{%
  \institution{Università degli Studi di Milano, Dipartimento di Informatica}
  \city{Milan}
  \country{Italy}
}

\author{Matteo Manzoni}
\email{matteo.manzoni@unimi.it}
\orcid{0009-0002-1946-5919}
\affiliation{%
  \institution{Università degli Studi di Milano, Dipartimento di Informatica}
  \city{Milan}
  \country{Italy}
}

\author{Filippo Corti}
\email{filippo.corti1@studenti.unimi.it}
\orcid{0009-0001-2493-6132}
\affiliation{%
  \institution{Università degli Studi di Milano, Dipartimento di Informatica}
  \city{Milan}
  \country{Italy}
}

\author{Sergio Mascetti}
\email{sergio.mascetti@unimi.it}
\orcid{0000-0002-8416-4023}
\affiliation{%
  \institution{Università degli Studi di Milano, Dipartimento di Informatica}
  \city{Milan}
  \country{Italy}
}

\renewcommand{\shortauthors}{Ahmetovic et al.}


\begin{abstract}
Shared control is a form of video gaming accessibility support that allows players with disabilities to delegate inaccessible controls to another person.
Through interviews involving 14 individuals with lived experience of accessible gaming in shared control, we explore the ways in which shared control technologies are adopted in practice, the accessibility challenges they address, and how the support currently provided in shared control can be automated to remove the need for a human assistant.
Findings indicate that shared control is essential for enabling access to otherwise inaccessible games, but its reliance on human support is a key limitation. Participants welcomed the idea of automating the support with software agents, while also identifying limitations and design requirements. Accordingly, this work contributes insights into current practices and proposes guidelines for developing automated support systems.
\end{abstract}



\maketitle

\section{Introduction}
\label{sec:introduction}
Video games have become a leading entertainment industry, with more than $3.35$ billion video game players worldwide~\cite{gamersWorldwide} ($41\%$ of the global population).
Among them, $20\%$ have some form of disability.
Furthermore, $46\%$ of people with disabilities report playing video games regularly~\cite{mosely2022video}.

Numerous assistive technologies have been developed to enable autonomous access to video games by players with disabilities~\cite{aguado2020accessibility}.
Despite these technologies, not all players are able to access every game independently~\cite{bierre2004accessibility,martinez2024playing}.
To address this issue, \emph{Shared Control} solutions have been developed, such as \textit{Xbox Controller Assist}~\cite{xboxshare}, allowing players with disabilities to delegate control of certain game actions to another person.
Recent research also explores how to extend this approach by replacing human support with a software agent~\cite{cimolino2021role}.
However, to the best of our knowledge, no prior study investigates how existing shared control technologies are commonly used by players, to address which accessibility challenges, and how these needs can be translated into technologies that automate such support.

To address these issues, we conduct a study, through interviews and focus groups, with $14$ individuals who have prior experience in the use of shared control systems, including people with disabilities who play using shared control, others who support players with disabilities during shared control, and experts in assistive gaming technologies.
Through reflexive thematic analysis, the research first aims to gain a deeper understanding of the shared control technologies currently in use, highlighting their benefits and limitations.
Second, we investigate the acceptability and the key features that a system automating the role of the supporting person should have.

Results show that shared control is essential for making certain games accessible that would otherwise not be playable.
However, these techniques have a major limitation: they require the availability of a person willing to provide support.
For this reason, participants welcomed the possibility of replacing the supporting person with a software agent, while at the same time highlighting both the potential limitations of the solution and the features it should possess.
Our work, therefore, provides a better understanding of the current use of shared control systems and outlines design guidelines for the development of shared control technologies with automated support.

\section{Background}
\label{sec:related}
This section surveys the accessibility problems faced by gamers with disabilities in playing video games (Section~\ref{ssec:related_accessibility}), and the existing assistive technologies designed to overcome these limitations (Section~\ref{ssec:related_technologies}).
Then an overview of the state of the art in shared control solutions is presented, both in general contexts (Section~\ref{ssec:related_sc}) and specifically when applied to gaming (Sections~\ref{ssub:related_sc_hc} and \ref{ssub:related_sc_pa}).

\subsection{Video Games Accessibility}
\label{ssec:related_accessibility}

Playing video games poses accessibility challenges for people with disabilities~\cite{martinez2024playing, bierre2005game, yuan2011game, aguado2020accessibility, gonccalves2023my, brown2021designing}.
These challenges include single-sense feedback (\textit{e.g.}, audio cues without visual counterparts)~\cite{brown2021designing, gonccalves2023my, martinez2024playing, bierre2005game}, unsuitable button layouts~\cite{martinez2024playing, dalgleish2023can, brown2021designing}, and demanding input gestures, such as repeated button mashing or long holds~\cite{bierre2005game, brown2021designing}.
Thus, after identifying a game that they find interesting, gamers with disabilities have to adapt the game to their specific needs~\cite{martinez2024playing}.
In some cases, however, adaptations fall short, forcing players to \quots{play their own game}, engaging it in unconventional ways~\cite{gonccalves2023my}, or even to abandon the game altogether~\cite{martinez2024playing}.
Guidelines written to facilitate the development of in-game accessibility features exist~\cite{bierre2004accessibility, martinez2024playing, bierre2005game, gonccalves2023my}.
However, these guidelines are rarely implemented~\cite{yuan2011game, aguado2020accessibility}.
Due to this, accessibility often needs to be achieved through outside-of-game solutions.
For example, participants in our interview often mentioned accessible game controllers and input-modifying software.

\subsection{Accessibility of Video Game Controls}
\label{ssec:related_technologies}

Various peripherals are used as video game controllers.
In particular, gamepads provided with major gaming consoles have evolved to share several common design features~\cite{lu2008evolution,maggiorini2017evolution}.
For example, Xbox and PlayStation controllers are designed to be held with both hands and require coordinated use of the thumbs (each capable of controlling a joystick and four buttons) as well as two additional fingers (usually the index and middle fingers) to operate two side buttons per hand.
This design assumes that all users have similar hand anatomy and size~\cite{brown2013evaluating}, can operate multiple inputs with both hands~\cite{dalgleish2023can}, and are able to perform quick, coordinated, and reactive movements~\cite{dalgleish2023can}.
As a result, standard controllers may be inaccessible to players with disabilities, for example, those with upper-limb mobility impairments.

Alternative controllers, designed to be more accessible and customizable, offer different button layouts from standard ones.
Some are designed to be operated with a single hand~\cite{xboxAdaptiveJoystick} or placed on a flat surface~\cite{liteSE,xboxAdaptiveController,ps5AccessController}.
Others are personalizable through external buttons~\cite{xboxAdaptiveController,flexController,ps5AccessController} or a modular design~\cite{proteusController}.

Software support has also been proposed to enhance the accessibility of game controllers.
Some tools allow remapping controller buttons to different commands; these are sometimes available directly within the games themselves.
Others allow remapping of commands across different types of input devices.
For example, JoyToKey~\cite{joyToKey} allows a game controller to emulate keyboard inputs for games designed to be played with a keyboard.
Further solutions enable users to perform actions using voice input~\cite{ahmetovic2021replay,voiceAttack} or facial expressions~\cite{manzoni2024personalized,playAbility}, which can then be mapped to specific game commands.

Through the use of these tools, inaccessible games can be made accessible to many people with disabilities.
The resulting game setup, that is the specific configuration of hardware and software accommodations and accessibility tools used, can be quite articulated, complex, and can vary based on the game, the disability, and the preferences of the player.
However, for some gamers with disabilities, existing accessibility tools may not suffice to access and effectively use all the inputs required to play a game~\cite{cimolino2021role}.
In such cases, one possibility is to delegate inaccessible game controls to someone else through shared control~\cite{cimolino2022two,xboxshare}.

\subsection{Shared Control}
\label{ssec:related_sc}

Shared control refers to two or more agents collaboratively interacting with a system.
Beyond video games, shared control has been studied for collaborative robot control~\cite{dragan2013policy, gopinath2016human}, assisted vehicle driving~\cite{li2018shared}, and management of shared user interfaces~\cite{dietz2001diamondtouch}.
In these contexts, the proposed interaction models involve either cooperation between humans or the integration of software agents to assist the user.

The terminology used to distinguish between different forms of support varies across application domains. 
For example, when referring to cooperation between only human actors, commonly used terms include 
\textit{human-human collaboration}~\cite{wang2020human}, 
\textit{multi-user interaction}~\cite{dietz2001diamondtouch}, 
and \textit{multi-operator-single-robot}~\cite{feth2009shared}.
Similarly, systems in which software agents contribute to the control are referred to using terms such as 
\textit{partial automation}~\cite{cimolino2022two}, 
\textit{human-AI collaboration}~\cite{wang2020human}, 
\textit{human-in-the-loop}~\cite{gopinath2016human}, 
or \textit{supervisory control}~\cite{backes1990umi}.

\begin{figure}[htb]
    \centering
    \includegraphics[width=0.7\linewidth]{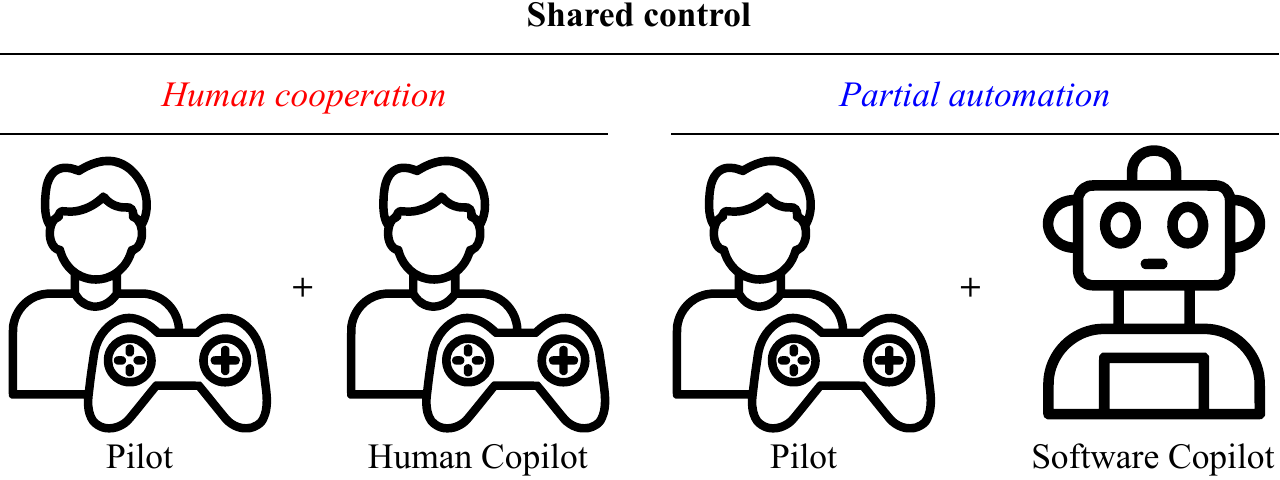}
    \caption{Relationship between shared control, human cooperation, and partial automation}
    \label{fig:scOverview}
    \Description{A diagram showing the relationship between shared control, human cooperation, and partial automation.}
\end{figure}

To disambiguate the various meanings, we use the term \textit{shared control} to refer to all interactions that allow game controls to be distributed across multiple actors, usually a \textit{pilot}, \textit{i.e.}, the person primarily responsible for playing (typically a person with disabilities), and a \textit{copilot} supporting the pilot.
As illustrated in Figure~\ref{fig:scOverview}, the two subtypes of shared control are: \textit{human cooperation}, when the copilot is a human actor; and \textit{partial automation}, when the copilot is a software agent.

\subsubsection{Human Cooperation for Video Games Accessibility}
\label{ssub:related_sc_hc}

Several human cooperation solutions have been proposed in the context of video games, although not explicitly as assistive technologies.
Instead, most studies have focused on analyzing how different forms of cooperation affect aspects such as social interaction, gameplay experience, and perceived enjoyment. 
Works by Loparev~\cite{loparev2014introducing}, Sykownik~\cite{sykownik2017exploring}, and Rozendaal~\cite{rozendaal2010exploring} observed that introducing interdependence mechanics between players led to higher levels of social interaction compared to independent play.
Conversely, Rozendaal et al.~\cite{rozendaal2010exploring} also noted that requiring cooperation to progress through the game could reduce players' perceived autonomy and control.

In the context of video game accessibility, Microsoft has released \textit{Xbox Controller Assist} (formerly \textit{Xbox Copilot})~\cite{xboxshare}, a software available on Xbox consoles and Windows PCs that allows linking two controllers so that two players can use them to control the same game actions, as if they were a single controller.
This solution is actively advertised as an accessibility feature \quots{to help a friend or loved one through a game or console experience}, and evidence of the usage of this solution by players with disabilities is available on various social media\footnote{A list of social network posts on this topic is available as supplemental material.}.
Similarly, on \textit{PlayStation 5} (PS5) consoles, the \textit{PS5 Access Controller}~\cite{ps5AccessController} can be paired with another PS5 controller, enabling human cooperation.
A third-party solution, the \textit{Titan Two} adapter~\cite{titanTwo}, supports remote human cooperation over the internet.
In this configuration, the game feedback (video and audio) is shared with a remote player who can play on their controller. Their input is then transmitted to the local console or PC, where it is merged with the local player's input, enabling collaborative control of the same game session even when players are not physically in the same location.

Gonçalves et al. have investigated human cooperation in relation to video game accessibility~\cite{gonccalves2021exploring}.
Specifically, they developed two video games intended to be played in pairs by individuals with mixed abilities: one sighted player would face a visual challenge, while one blind player would face an auditory one.
Their findings demonstrate that cooperation between players with differing abilities is not only possible but can be effectively supported through game design that leverages each participant's strengths.
However, the authors themselves acknowledge that this form of cooperation does not constitute a generalizable accessibility solution, as it relies on custom-designed games and does not apply to disabilities beyond visual impairments.

\subsubsection{Partial Automation for Video Games Accessibility}
\label{ssub:related_sc_pa}

Partial automation is a type of shared control that removes the necessity for human copilots, substituting them with a software agent.
In the video game accessibility domain, partial automation has been implemented in commercial video games~\cite{callOfDuty,forzaMotorsportAccessibility,bayonetta,bayonetta2,marioKart,kingdomComeDeliverance}, and it has been explored in prior research as well~\cite{cimolino2021role,cimolino2023automation,hougaard2021willed,vicencio2014effectiveness,hwang2017game,cechanowicz2014improving}.
However, the application of partial automation has been mostly limited to specific game tasks.
One use case is player balancing: providing assistance in a given task that is inversely proportional to the player's skill level, thus ensuring a similar gaming experience for all.
For example, some first-person shooters such as \textit{Call of Duty}~\cite{callOfDuty} integrate aim-assistance, whose effectiveness has been evaluated in prior research~\cite{vicencio2014effectiveness,hwang2017game}.
Similarly, some racing games implement steering assistance to help players stay on the road~\cite{cechanowicz2014improving,marioKart}.
Another approach is to reduce the effort required to play by automating some inputs.
For example, some racing games such as \textit{Forza Motorsport}~\cite{forzaMotorsportAccessibility} implement automatic gear shifting. 
More pronounced implementations of partial automation are found in games like the \textit{Bayonetta} series~\cite{bayonetta,bayonetta2}, which feature an automatic mode where the player only needs to focus on attacking while the game autonomously handles movement and enemy targeting.
Another example is \textit{Zac - O Esquilo}~\cite{medeiros2015developing}, a one-switch (controlled with only one button) video game in which the player decides when to move the avatar, but an algorithm determines the direction of the movement.
These solutions are specifically designed and implemented for each game, but lack generalizability, which is a significant barrier to the widespread adoption of partial automation as a solution for video game accessibility.

Cimolino et al.~\cite{cimolino2021role} employ partial automation in two research video games to demonstrate how distributing gameplay actions between a player and a software agent can enable players with motor disabilities to play.
In this approach, the player performs only the actions that they are physically capable of, while the remaining actions are delegated to the software agent.
Findings from the study highlight the importance of mutual understanding between the player and the agent.
Lack of mutual understanding may lead to \textit{automation confusion}~\cite{cimolino2023automation}, a phenomenon in which the player struggles to distinguish the outcome of their own actions from those performed by the agent.

These works contribute to a better understanding of how partial automation can support video game accessibility.
However, to the best of our knowledge, no prior work explores existing shared control practices, how they are used, and with what purposes.
Most importantly, it is unclear how the support currently provided through human cooperation can be provided through generalizable partial automation solutions and how such solutions should be designed.

\section{Methodology}
\label{sec:methodology}
To understand how human cooperation technologies are currently used by people with disabilities, and to inform the design of future partial automation approaches, semi-structured interviews and focus groups~\cite{lazar2017research} were organized with people who regularly use, or have used in the past, human cooperation systems for video games.
The experimentation was approved by the Ethics Committee of \anon[our institution]{the University of Milan (opinion no. 16/25)}.

\subsection{Participants}
\label{sec:participants}

Participants were recruited through announcements on social media pages frequented by people with disabilities, such as \textit{r\slash disabledgamers} on Reddit, and by reaching out to accessibility experts' organizations.
A local association for people with motor impairments, \anon{Spazio Vita\footnote{\url{https://spaziovitaniguarda.it}} of Niguarda Hospital in Milan - Italy}, 
also supported the recruitment by mediating contact with some participants.
The following inclusion criteria were adopted:
\begin{itemize}
    \item Being of legal age.
    \item Being able to speak \anon[languages spoken by the authors]{Italian or English}.
    \item One of the following:
        \begin{itemize}
        \item (\textbf{Pilot}) Using or having used human cooperation in the past.
        \item (\textbf{Copilot}) Using or having used human cooperation in the past to support a person with disabilities.
        \item (\textbf{Expert}) Being an accessibility expert who uses human cooperation as a tool to help people with disabilities in playing video games.
    \end{itemize}
\end{itemize}

We recruited 14 participants, among whom 8 were pilots, 5 copilots, and 6 experts.
Participants' full demographic profiles are reported in Table~\ref{tab:demographicData}.
For organizational reasons, it was not possible to collect demographic data for \ParticipantNine{}.
One additional participant was excluded from the results because they used Xbox Controller Assist~\cite{xboxshare} not in human cooperation, but to play alone with two controllers.
In total, 4 individual interviews and 4 focus groups were conducted:
\begin{itemize}
    \item Individual interviews:
    \begin{itemize}
        \item \ParticipantOne{}: tried but no longer uses human cooperation, having found a gaming setup for playing independently.
        \item \ParticipantTwo{}: uses remote human cooperation to play with people met online.
        \item \ParticipantSix{}: accessibility professional with experience in hardware solutions for accessible gaming.
        \item \ParticipantSeven{}: accessibility expert with occasional gaming experience.
    \end{itemize}

    \item Focus groups:
    \begin{itemize}
        \item \ParticipantFour{} and \ParticipantFive{}: \ParticipantFive{} is a family member, caregiver, and habitual copilot of \ParticipantFour{}.
        
        \item \ParticipantEight{}, \ParticipantNine{}, \ParticipantTen{}, and \ParticipantEleven{}: members of a weekly gaming group in which they play cooperatively. \ParticipantEight{} usually pairs with \ParticipantNine{}, while \ParticipantTen{} plays with \ParticipantEleven{}.
        
        \item \ParticipantTwelve{} and \ParticipantThirteen{}: coordinators of the gaming group involving \ParticipantEight{}, \ParticipantNine{}, \ParticipantTen{} and \ParticipantEleven{}. They support the participants both as copilots and by helping to find suitable game setups.
        
        \item \ParticipantFourteen{} and \ParticipantFifteen{}: colleagues who assist people with disabilities in identifying and configuring accessible game setups.
    \end{itemize}
\end{itemize}

\begin{table}[htb]
\caption{Participants' Demographic Data. The participant's role is in ID's subscript: P - Pilot, C - Copilot, E - Expert}
\label{tab:demographicData}
\centering
\begin{tabular}{|l|c|c|c|c|c|c|c|}
\hline
\multirow{2}{*}{\textbf{ID}} & \multirow{2}{*}{\textbf{Age}} & \multirow{2}{*}{\textbf{Gender}} & \textbf{Gameplay} & \multicolumn{2}{c|}{\textbf{Disability}} & \textbf{Difficulties} & \textbf{Preferred} \\ \cline{5-6}
& & &  \textbf{frequency} & \textbf{Type} & \textbf{Onset} & \textbf{in gameplay} & \textbf{platform} \\ \hline
\ParticipantOne{} & 29-38 & M & Daily & {Paresis} & {Birth} & Moderately & Switch, PC, PS \\ \hline
\ParticipantTwo{} & 29-38 & M & Daily & {Blindness} & {Birth} & A lot & PC, PS \\ \hline
\ParticipantFour{} & 29-38 & F & Daily & Spastic quadriplegia & {Birth} & A lot & Xbox \\ \hline
\ParticipantFive{} & 39-50 & M & Daily & None & None & None & Smartphone, Xbox \\ \hline
\ParticipantSix{} & 39-50 & M & Daily & {Burn injury} & {20 years} & A little & Xbox \\ \hline
\ParticipantSeven{} & 29-38 & M & Monthly & None & None & None & Xbox \\ \hline
\ParticipantEight{} & 18-28 & M & Daily & Spastic tetraparesis & {Birth} & Moderately & Switch, PS5 \\ \hline
\ParticipantNine{} & \multicolumn{3}{c|}{Data Not Available} & {Yes} & \multicolumn{3}{c|}{Data Not Available} \\ \hline
\ParticipantTen{} & 39-50 & F & Daily & Pediatric tetraparesis & {Birth} & Moderately & Smartphone \\ \hline
\ParticipantEleven{} & 18-28 & F & Daily & Reduced arm mobility & {Birth} & A little & Tablet \\ \hline
\ParticipantTwelve{} & 39-50 & M & Weekly & None & None & None & PC, Xbox \\ \hline
\ParticipantThirteen{} & 39-50 & M & Weekly & None & None & None & PC \\ \hline
\ParticipantFourteen{} & 29-38 & F & Monthly & None & None & None & Switch \\ \hline
\ParticipantFifteen{} & 39-50 & M & Weekly & None & None & None & Switch, Xbox \\ \hline
\end{tabular}%
\end{table}

Ten participants identified as male, while four identified as female.
The most represented age groups were 29--38 and 39--50 (6 participants each), followed by 18--28 age group (2 participants).
Eight participants declared having a disability, among which \ParticipantTwo{} is the only one with a visual impairment, while the others have motor disabilities.
The remaining six participants do not have disabilities, but have experience in supporting people with disabilities in using human cooperation.
In particular, \ParticipantSix{}, \ParticipantSeven{}, \ParticipantTwelve{}, \ParticipantThirteen{}, \ParticipantFourteen{}, and \ParticipantFifteen{} professionally help people with disabilities in finding hardware and software configurations for gaming according to their needs.

All participants play video games at least once a month.
Among those with disabilities, only \ParticipantSix{} and \ParticipantEleven{} reported having little difficulty in playing video games; the others declared having at least moderate difficulty.
The most used gaming platform is Xbox (6 participants), followed by PC and Nintendo Switch (4 participants each), PlayStation (3 participants), smartphone (2 participants each), and tablet (1 participant).
Nine people attribute their reference to the ease of access of the platform (\ParticipantOne{}, \ParticipantTwo{}, \ParticipantFour{}, \ParticipantFive{}, \ParticipantSeven{}, \ParticipantTen{}, \ParticipantEleven{}, \ParticipantTwelve{}, \ParticipantFifteen{}).
Other motivations include familiarity with it (\ParticipantFive{}, \ParticipantSix{}, \ParticipantThirteen{}) and the presence of exclusive video games for the platform (\ParticipantEight{}, \ParticipantFourteen{}).
Eight of the nine participants with disabilities also use hardware or software tools to facilitate access to video games.
These include the Xbox Adaptive Controller (\ParticipantFour{}, \ParticipantEight{}, \ParticipantNine{}, \ParticipantTen{}, \ParticipantEleven{}), custom buttons and joysticks (\ParticipantEight{}, \ParticipantTen{}, \ParticipantEleven{}) and software for remapping controller and keyboard keys (\ParticipantOne{}, \ParticipantNine{}).

\subsection{Procedure}
\label{sec:procedure}

The interviews were conducted online, using video-calling platforms, and were audio-recorded to facilitate the following analysis.
In addition to the participant, at least two members of the research group were always present during each interview: one with the role of main interviewer and the other with a support function, tasked with intervening with additional questions not initially planned.
Each interview lasted between 60 and 90 minutes.
In cases where users knew each other and shared usage modalities of human cooperation systems, the interviews were organized as focus groups, in which multiple users participated in the meeting and also discussed among themselves on the proposed topics.

The initial interview outline\footnote{The initial interview outline is available as supplemental material.} drew inspiration from prior works on shared control and partial automation~\cite{cimolino2022two, cimolino2022impact, cimolino2023automation, gonccalves2021exploring, rozendaal2010exploring}.
%
The first part of each interview was dedicated to human cooperation technologies.
With pilots and copilots, we explored how they used these technologies, which ones they preferred, and what advantages and limitations they had.
With accessibility experts, we investigated how they support users who come seeking a new gaming setup tailored to their needs.
We then proceeded to understand how they integrate human cooperation technologies into these setups.

In the second part, a concept of a partial automation system for video game accessibility was presented, and interviewees were asked for their opinion, how they would use the system, and what requirements it should have.
The system was described as a shared control solution, similar to Xbox Controller Assist~\cite{xboxshare}, but designed to allow playing without the need for a second person.
We indicated that the users would be able to select which actions they wanted to control, while the system would autonomously control the remaining ones.
The description of the system was purposefully kept vague to encourage the elicitation of its potential functionalities.
The initial questions were expanded based on the outcome of the first two interviews.
In particular, two additional topics on human cooperation were introduced: communication with the copilot during gameplay and reasons for not using this technology.

\subsection{Data Analysis}
\label{sec:analysis}

The audio-recordings of the interviews were transcribed and analyzed following the reflexive thematic analysis methodology~\cite{terry2017thematic}.
The first interview was analyzed jointly by four researchers, identifying initial codes.
Subsequent interviews were randomly divided among the researchers.
Each interview was coded by one researcher and subsequently reviewed by a second researcher, who integrated any additional observations.
In a subsequent meeting, all codes were reviewed and discussed as a group to resolve any ambiguities or disagreements, re-examining the relevant parts of the transcriptions until consensus was reached.
Finally, the extracted codes were consolidated and organized into sub-themes and main themes through an iterative process of comparison among researchers.

\section{Results}
\label{sec:results}
Six main themes were identified (Table~\ref{tab:themes}):
\textit{Shared Control Benefits and Limitations} (Section~\ref{ssec:sc-benefits}),
\textit{Human Cooperation Limitations Addressed in Partial Automation} (Section~\ref{ssec:sc-limitations}),
\textit{Copilot's Interventions} (Section~\ref{ssec:copilot-interventions}),
\textit{Negotiating Collaboration} (Section~\ref{ssec:negotiating-collaboration}),
\textit{Interaction} (Section~\ref{ssec:interaction}), and
\textit{Factors Affecting the Collaboration} (Section~\ref{ssec:factors-affecting-collaboration}).
We observe that, although the interviews focused on two distinct areas — human cooperation technologies and the proposal of a partial automation system — we unify their thematic analysis, treating the two areas orthogonally with respect to the identified themes when possible. This choice is motivated by the similarity of the collected codes, which are largely transversal to both areas.

\begin{table}[htb]
\caption{Themes and sub-themes identified through the reflexive thematic analysis}
\label{tab:themes}
\centering
\begin{tabular}{|l|l|}
\hline
\textbf{Themes} & \textbf{Sub-themes} \\ \hline

\multirow{3}{*}{\hyperref[ssec:sc-benefits]{1. \textit{Shared Control Benefits and Limitations}}} 
    & \hyperref[ssec:sc-benefits-accessibility]{1. \textit{Accessibility}} \\
    & \hyperref[ssec:sc-benefits-sociality]{2. \textit{Sociality and Feeling of Inclusion}} \\
    & \hyperref[ssec:sc-limitations-ethical]{3. \textit{Ethical Concerns}} \\
    \hline

\multirow{3}{*}{\hyperref[ssec:sc-limitations]{2. \textit{Human Cooperation Limitations Addressed in Partial Automation}}} 
    & \hyperref[ssec:sc-limitations-tradeoff]{1. \textit{Loss of Autonomy}} \\
    & \hyperref[ssec:sc-limitations-availability]{2. \textit{Copilot Availability}} \\
    & \hyperref[ssec:sc-limitations-engagement]{3. \textit{Copilot's Engagement During Play}} \\
    \hline

\multirow{5}{*}{\hyperref[ssec:copilot-interventions]{3. \textit{Copilot's Interventions}}} 
    & \hyperref[ssec:copilot-interventions-during-setup]{1. \textit{Assistance During Game Setup}} \\
    & \hyperref[ssec:copilot-interventions-menu-access]{2. \textit{Assistance with Menu Access}} \\
    & \hyperref[ssec:copilot-interventions-playing]{3. \textit{Assistance by Playing}} \\
    & \hyperref[ssec:copilot-interventions-signaling]{4. \textit{Assistance by Signaling}} \\
    & \hyperref[ssec:copilot-interventions-suggesting]{5. \textit{Assistance by Suggesting}} \\ \hline

\multirow{4}{*}{\hyperref[ssec:negotiating-collaboration]{4. \textit{Negotiating Collaboration}}} 
    & \hyperref[ssec:negotiating-collaboration-actions]{1. \textit{Actions Separation}} \\
    & \hyperref[ssec:negotiating-collaboration-policies]{2. \textit{Policies Guiding Action Assignment}} \\
    & \hyperref[ssec:negotiating-collaboration-leadership]{3. \textit{Leadership Management}} \\
    & \hyperref[ssec:negotiating-collaboration-copilot]{4. \textit{Copilot's Operational Autonomy}} \\ \hline

\multirow{2}{*}{\hyperref[ssec:interaction]{5. \textit{Interaction}}} 
    & \hyperref[ssec:interaction-communication]{1. \textit{Verbal and Non-Verbal Communication}} \\
    & \hyperref[ssec:interaction-intent]{2. \textit{Intent Understanding}} \\ \hline

\multirow{2}{*}{\hyperref[ssec:factors-affecting-collaboration]{6. \textit{Factors Affecting the Collaboration}}} 
    & \hyperref[ssec:factors-affecting-collaboration-knowledge]{1. \textit{Knowledge of the Game}} \\
    & \hyperref[ssec:factors-affecting-collaboration-relation]{2. \textit{Relationship Between Pilot and Copilot}} \\ \hline

\end{tabular}
\Description{The table presents the themes and sub-themes identified through reflexive thematic analysis. 
It is organized in two columns: the left column lists the main themes, 
while the right column specifies the corresponding sub-themes associated with each theme.}
\end{table}

\subsection{Shared Control Benefits and Limitations}
\label{ssec:sc-benefits}

This section explores the benefits of the shared control technology as an accessibility tool (Section~\ref{ssec:sc-benefits-accessibility}), its impact on sociality and inclusion (Section~\ref{ssec:sc-benefits-sociality}), and ethical concerns related to the misuse of this technology (Section~\ref{ssec:sc-limitations-ethical}).

\subsubsection{Accessibility}
\label{ssec:sc-benefits-accessibility}


\hc{}
This approach enables people with disabilities to access potentially any game (\ParticipantTwo{}) that is not accessible using other assistive technologies (\ParticipantTwo{}, \ParticipantFour{}, \ParticipantEight{}, \ParticipantNine{}, \ParticipantTen{}, \ParticipantEleven{}).
It is also beneficial for those games that can be played autonomously, but at a cost of a high physical or cognitive load (\ParticipantEight{}).
\ParticipantSix{}, \ParticipantSeven{}, \ParticipantFourteen{}, and \ParticipantFifteen{}, who provide support to people with disabilities in setting up games and devising accessible game configurations, confirm that human cooperation is indeed useful as an assistive solution, albeit with some limitations (Section~\ref{ssec:sc-limitations}).
As observed by \ParticipantSix{}, playing with human cooperation can be initially frustrating, but it may be the only solution to play certain games.
\begin{displayquote}
\ParticipantSix{}: \it
The psychosocial effect is that there is going to be frustration, but there's also going to be a lot of joy. So you either can't play at all, or you play with a copilot, and you guys learn and grow together.
\end{displayquote}

\pa{}
The idea of introducing partial automation as an alternative to human cooperation was acclaimed by most participants, who recognize its potential as an assistive technology.
\ParticipantOne{} considers that, similarly to human cooperation, partial automation would be useful to access games too complex to be played autonomously.
\ParticipantTwo{}, \ParticipantEight{}, \ParticipantNine{},\ParticipantTen{}, and \ParticipantEleven{} furthermore highlight that partial automation could be particularly useful in multiplayer games, which commonly have fewer accessibility options.
\ParticipantEight{} and \ParticipantEleven{} would be eager to test this solution.
For \ParticipantSix{}, partial automation could also be effective for users with very low mobility: in such cases, the user could delegate most interactions to the system, while still preserving the decisional role on the progress of the game. That would make the gaming experience more similar to an interactive movie, preserving the narrative content and the active role of the player.


\subsubsection{Sociality and Feeling of Inclusion}
\label{ssec:sc-benefits-sociality}

\hc{}
Many participants pointed out that human cooperation is not simply about being able to play, but it's also about socializing through playing together. For this reason, \ParticipantSix{} strongly advocates for human cooperation, even when other accessibility options are available.
\begin{displayquote}
\ParticipantSix{}: \it
And so, yeah, I think that there is a lot of good psychosocial component to playing together. And when you play alone, it's ok... It's just I think it's better with other people.
\end{displayquote}
\ParticipantFifteen{} noted that playing through human cooperation can also strengthen the bond with the copilot, for instance, when the latter is a family member.
That's why, when working with children, \ParticipantFifteen{} usually recommends human cooperation as a solution. 
Furthermore, \ParticipantEight{}, \ParticipantNine{}, \ParticipantTen{}, and \ParticipantEleven{}, who participate together in weekly gaming workshops (Section~\ref{sec:participants}), reported that these meetings are also an opportunity to socialize.
\ParticipantFour{} and \ParticipantFifteen{} further explained that human cooperation enables them to discuss with friends about games they would otherwise be unable to access.
Without human cooperation, \ParticipantFour{} would be limited to watching others play, which could lead to feelings of exclusion.
According to \ParticipantFive{}, human cooperation can also be used to introduce inexperienced players to gaming.
For example, allowing a parent to learn to play alongside their children, thus fostering family bonding.  

\pa{}
In the absence of a human copilot, the social dimension of partial automation may be reduced.
For this reason, \ParticipantFour{} would still prefer to play with \ParticipantFive{} via human cooperation, even if a partial automation system were available, as those moments provide an opportunity to spend time together.
\ParticipantEight{} and \ParticipantThirteen{} agree on this aspect, expressing that they would not want to abandon human cooperation entirely to avoid losing the social benefits it entails.
In contrast, \ParticipantOne{}, \ParticipantTwelve{}, and \ParticipantFifteen{} observed that partial automation does not necessarily remove the social dimension; rather, it may create new opportunities.
Indeed, partial automation could enable players to independently access multiplayer games, which by their nature foster social interaction and group play.
Similar to human cooperation, partial automation could also facilitate the inclusion of new players in gaming.


\subsubsection{Ethical Concerns}
\label{ssec:sc-limitations-ethical}

Some accessibility solutions can be misused or exploited for cheating, undermining the intended gameplay and negatively affecting other players' experiences.
For this reason, multiplayer games may restrict their use.

\hc{} 
\ParticipantFive{}, \ParticipantSix{}, and \ParticipantFifteen{} note that human cooperation systems are subject to this issue. As \ParticipantTwo{} points out, the problem is exacerbated by the fact that multiplayer games are generally less accessible than their single-player counterparts and would therefore need better accessibility support.

\pa{}
A partial automation system would face similar limitations. For this reason, \ParticipantSeven{} is skeptical about its applicability in multiplayer contexts. \ParticipantTwo{} suggests distinguishing between competitive and non-competitive multiplayer games: while in the former, the use of partial automation should be regulated, in the latter it should remain unrestricted, as should be the case for single-player games. To address this concern, \ParticipantThirteen{} proposes the development of partial automation systems capable of learning and adapting to the player’s abilities, thereby moderating the level of assistance provided. This aspect is discussed in more detail in Section~\ref{ssec:copilot-interventions-playing}. Overall, \ParticipantSix{} expects mixed reactions:
\begin{displayquote}
\ParticipantSix{}: \it
The people who need it would celebrate, the people who don't need it would think it's cheating. I mean, some of them, not all of them. I am able to play, but I would celebrate because I know that so many of the people that I've worked with will now be able to play more. 
\end{displayquote}

\subsection{Human Cooperation Limitations Addressed in Partial Automation}
\label{ssec:sc-limitations}

Human cooperation also has limitations that affect its widespread adoption: 
reduced pilot's autonomy (Section~\ref{ssec:sc-limitations-tradeoff}),
need for a copilot (Section~\ref{ssec:sc-limitations-availability}), 
and concerns about the copilot’s engagement during play (Section~\ref{ssec:sc-limitations-engagement}).

\subsubsection{Loss of Autonomy}
\label{ssec:sc-limitations-tradeoff}

\hc{}
Human cooperation implies dependence on another person, which limits the player's autonomy.
Participants react to this limitation in different ways.
For \ParticipantTwo{}, who is blind, no other accessibility solution is viable.
While not uncomfortable requesting help, \ParticipantTwo{} acknowledges that dependence can limit play options.
In contrast, \ParticipantOne{} avoids human cooperation despite other tools being less effective.
This is in part due to logistical barriers, and in part because having to ask for help every time would feel burdensome.
\ParticipantFourteen{} confirms this reluctance, noting that it is particularly common among people with acquired disabilities, possibly due to the inability to regain pre-disability skill levels.
%
\begin{displayquote}
\ParticipantFourteen{}: \it
There will always be people who, if they can't have full access to the game they want to be able to play, would choose not to play. We have definitely met people that would say: \quots{If I can't do this myself completely independently, I'd rather not do it.}. [...] I think most people would want to be able to play the way that they played before, and so it might not be as enjoyable if they don't have full access.
\end{displayquote}
For some, relying on other people's assistance should never be an option.
\ParticipantTwo{} reports getting criticized by other players with visual impairments for relying on assistance from another person, believing that this misrepresents the gaming experience of a visually impaired player.
In general, \ParticipantSix{}, \ParticipantSeven{}, \ParticipantFourteen{}, and \ParticipantFifteen{} would prioritize finding game configurations that preserve independence, without the necessity to rely on human cooperation.

\pa{}
Partial automation eliminates the dependency on another person, offering greater autonomy to the player.
For example, if such a solution were available, \ParticipantTwo{} would attempt higher difficulty levels, replay games multiple times, and commit to longer sessions without the feeling of weighing on someone else.
For this reason, \ParticipantSeven{} explained that, if partial automation were available, it would be a \textit{second-line option}, adopted when no complete configuration based on traditional accessibility solutions can be identified. 
Human cooperation would therefore be considered only a \textit{third-line option}, used solely in the absence of partial automation. 
Similarly, \ParticipantNine{} and \ParticipantTen{} highlighted that the independence afforded by partial automation constitutes a clear advantage over human cooperation approaches.


\subsubsection{Copilot Availability}
\label{ssec:sc-limitations-availability}

\hc{}
Human cooperation is also limited by the copilot's availability.
One side of the problem is logistical, since
the pilot and the copilot need to be physically co-located to play.
All participants reported that this requirement is a significant constraint in the use of human cooperation, as it drastically reduces the pool of potential copilots.
A common choice is to have a family member as the copilot, which is what \ParticipantFour{} and \ParticipantFive{} do.
The relationship between the pilot and the copilot is discussed in Section~\ref{ssec:factors-affecting-collaboration-relation}.
Additionally, the copilot may not always have time to play (\ParticipantFour{}).
To overcome the challenges of copilot's availability, some participants arrange gaming sessions in shared physical spaces.
For \ParticipantEight{}, \ParticipantNine{}, \ParticipantTen{}, and \ParticipantEleven{}, who meet at weekly gaming workshops, finding a copilot is straightforward, but only during scheduled sessions.
An alternative, known only to \ParticipantTwo{}, is remote human cooperation (Section~\ref{ssub:related_sc_hc}).
However, this approach requires a complex configuration of third-party hardware and software, and it introduces interaction latency that complicates the coordination.
\ParticipantTwo{} notes that such a solution, if natively integrated in gaming platforms, would make it easy to meet copilots, thus improving game accessibility for many people with disabilities.
When no copilot is available, fallback strategies vary.
\ParticipantTwo{}, \ParticipantNine{}, and \ParticipantEleven{} stop playing altogether; \ParticipantFour{}, \ParticipantEight{}, and \ParticipantTen{} switch to more accessible alternatives, such as mobile games.
Finally, \ParticipantTwo{}, \ParticipantTen{}, and \ParticipantEleven{} sometimes play with different copilots, which introduces the challenge of learning to play together (Section~\ref{ssec:factors-affecting-collaboration-relation}).

\pa{}
Partial automation doesn't have copilot availability issues.
That's why \ParticipantFour{} would find a system based on it \quots{fantastic} as it would allow playing even when no one is available as a copilot.
For \ParticipantTwo{}, this approach would benefit the copilots as well, by freeing them from the time commitment of learning and playing the entire game.
\begin{displayquote}
\ParticipantTwo{}: \it
The benefit of it is that you don't have to then wait for a person to be available, you can play in theory any game at any time [...] you could play on your own terms without having to then rope a second person in and make them take hundreds of hours of their life.
\end{displayquote}


\subsubsection{Copilot's Engagement During Play}
\label{ssec:sc-limitations-engagement}

\hc{}
\ParticipantOne{} notes that the gaming experience in human cooperation is not fully comparable to that offered by collaborative multiplayer games.
In the latter, the roles of the two players are generally equivalent or equally significant, whereas in human cooperation, the division of tasks is often unbalanced.
According to \ParticipantOne{}, this imbalance can affect the copilot’s enjoyment, as they are frequently assigned only a supporting role.
Indeed, \ParticipantOne{} considers playing as a copilot to be a predominantly negative experience, describing it as an activity performed mainly to help someone else rather than for personal enjoyment.
It should be noted, however, that \ParticipantOne{} has always acted as a pilot and has never directly experienced the role of copilot.
In contrast, \ParticipantFive{} reported finding their experience as a copilot rewarding, viewing it also as an opportunity to discover new video games.

\pa{}
With partial automation, this issue does not arise, as the copilot’s role is replaced by a software.



\subsection{Copilot's Interventions}
\label{ssec:copilot-interventions}

The copilot can support the pilot during gameplay in various ways: 
assistance with the setup before starting to play (Section~\ref{ssec:copilot-interventions-during-setup}), support with menu access (Section~\ref{ssec:copilot-interventions-menu-access}),
direct control of commands during gameplay (Section~\ref{ssec:copilot-interventions-playing}), signaling elements of interest in the game (Section~\ref{ssec:copilot-interventions-signaling}),
and suggestions on how to proceed in the game (Section~\ref{ssec:copilot-interventions-suggesting}).

\subsubsection{Assistance During Game Setup}
\label{ssec:copilot-interventions-during-setup}

Setting up a game configuration is a long and complex process, which in some cases can lead to frustration (\ParticipantFourteen{}) or even to giving up playing (\ParticipantTwelve{}), as also noted in prior literature~\cite{martinez2024playing}. The presence of a copilot can facilitate these operations.

\hc{}
Some participants are only able to play if another person helps them to prepare the gaming hardware and software setup (\ParticipantFour{}, \ParticipantNine{}, \ParticipantTen{}, \ParticipantEleven{}, \ParticipantFourteen{}, and\ParticipantFifteen{}).
Support can sometimes be needed even for those who can afterward play independently (\ParticipantSix{}, \ParticipantSeven{}, \ParticipantTwelve{}).
For example, the player may be able to play independently using an accessible controller, but may need support to connect, position, and configure the controller itself.
\ParticipantSix{} and \ParticipantSeven{} always try to maximize the user's autonomy, but some users still request their help in these situations.

\pa{}
While none of the participants commented about the (lack of) assistance during setup in a partial automation solution, we note that a software copilot cannot physically assist the pilot in preparing the gaming peripherals' configuration before starting to play.
Thus, in this case, human support may still be required.


\subsubsection{Assistance with Menu Access}
\label{ssec:copilot-interventions-menu-access}

Enabling in-game accessibility options through the game menu is sometimes needed to make a game accessible (\ParticipantTwo{}, \ParticipantSix{}, \ParticipantSeven{}, \ParticipantTwelve{}, \ParticipantThirteen{}, \ParticipantFourteen{}).
However, the menus themselves may lack accessibility (\ParticipantTwo{}, \ParticipantTwelve{}, \ParticipantThirteen{}), and solutions used to make the active gameplay accessible often do not work for accessing menus (\ParticipantThirteen{}).
To access text from game menus and navigate them \ParticipantTwo{} uses Optical Character Recognition tools. This introduces additional workload and, as \ParticipantTwo{} points out, these technologies often make mistakes.

\hc{}
Due to the above problem, \ParticipantTwo{} highlights that having a copilot ready to help with menu access is much preferable.
\ParticipantTwelve{} and \ParticipantThirteen{} also point out the issue of menu accessibility, explaining that people participating in their workshop very rarely manage to start a gaming session without help from a copilot.
%
\begin{displayquote}
\ParticipantThirteen{}: \it
If games were designed with accessibility in mind from the start, with continuous scrolling menus where you only need to click on the options, that would be one thing. But menu accessibility is rarely implemented, even in accessible games. It may seem trivial, but it's extremely limiting for our user base.
\end{displayquote}

\pa{}
\ParticipantTwelve{} and \ParticipantThirteen{} highlight that partial automation could support menu navigation.
By doing so, the system would allow the pilot to start a game by themselves, removing the need for other accessibility tools.


\subsubsection{Assistance by Playing}
\label{ssec:copilot-interventions-playing}

\hc{}
Assistance by controlling some of the game actions is usually the main form of support provided by the copilot.
\ParticipantOne{}, \ParticipantFour{}, \ParticipantEight{}, and \ParticipantNine{} all play with a copilot who primarily assists them by controlling actions in the game.

\pa{}
All participants envisioned partial automation assisting them by directly controlling some of the game actions, similarly to a human copilot.
However, some expressed doubts about the copilot’s ability to balance the level of support provided, avoiding being too skilled compared to the pilot and thus undermining the gameplay experience (\ParticipantOne{}, \ParticipantSix{}, \ParticipantTwelve{}, and \ParticipantThirteen{}).
In this sense, \ParticipantTwelve{} and \ParticipantThirteen{} expect partial automation to bring the player to \quots{the same level as others}, prioritizing the pilot's enjoyment over in-game success.
According to \ParticipantSix{}, the main goal should be to enable the pilot to reach a minimum standard that ensures enjoyment, without completely replacing their skill.
\ParticipantTwelve{} and \ParticipantThirteen{} also suggested that the level of support could be dynamically adjusted, progressively decreasing as the pilot improves their performance.
If this is not possible, \ParticipantFive{} and \ParticipantTwelve{} would at least like to manually select the desired level of support.


\subsubsection{Assistance by Signaling}
\label{ssec:copilot-interventions-signaling}

\hc{}
Another way of assisting the pilot is by signaling what's on the screen and what is happening in the game. This support is particularly valuable to \ParticipantTwo{}, who is blind. Indeed, \ParticipantTwo{} reports that their copilot must clearly indicate what is happening and which actions they are performing, so that they can coordinate effectively.

\pa{}
\ParticipantTwo{} states that partial automation should also provide this type of support, describing what is happening in the game rather than giving instructions on what to do. This way, the player can make their own decisions on how to act.
\begin{displayquote}
\ParticipantTwo{}: \it
If you give me that information, I can do what most players would be able to do. I'm going to decide who I target first. I'm going to decide how I play this. Whereas if you're just controlling me, I'm like \quots{Yeah, I can shoot when you tell me to shoot, but where's the fun in that?}.
\end{displayquote}


\subsubsection{Assistance by Suggesting}
\label{ssec:copilot-interventions-suggesting}
Finally, the copilot may support the pilot by providing suggestions.
This is particularly useful when the pilot is struggling to make progress in the game.

\hc{}
\ParticipantTwelve{} and \ParticipantThirteen{} provide this type of support 
in their weekly gaming group.
However, both note that, when giving suggestions, they always try to guide the pilot towards reasoning, without fully solving their problem.
\begin{displayquote}
\ParticipantTwelve{}: \it
I tell them \quots{Let's go back, maybe we missed something} and when we go back they notice something on their own that they didn't see earlier, like a handle or something to pull. Then they can go on by themselves.
\end{displayquote}

\pa{}
In partial automation \ParticipantTwo{} would like a similar support, where the copilot does not intervene in the game but provides suggestions.
Indeed, while it is possible for the copilot taking control of the game to resolve a situation unclear to the pilot, \ParticipantTwo{} notes that it is preferable not to intervene directly to preserve the pilot's sense of agency.
\ParticipantTwelve{} also imagines that the system could detect when the player is stuck in the game and give them suggestions to help them understand how to proceed.
For \ParticipantTen{} and \ParticipantTwelve{}, it is important that suggestions do not directly reveal the solution, so as to leave the player with the satisfaction of solving the puzzle themselves.
Finally, some participants suggest that the copilot could initiate communication proactively, for example, by recognizing moments of difficulty and spontaneously offering their help (\ParticipantOne{}, \ParticipantTwo{}, \ParticipantThirteen{}).

\subsection{Negotiating Collaboration}
\label{ssec:negotiating-collaboration}

Participants discussed the distribution of the game actions between the pilot and the copilot (Section~\ref{ssec:negotiating-collaboration-actions}), the criteria for the action assignment (Section~\ref{ssec:negotiating-collaboration-policies}), and leadership in the decision-making.
For this last aspect, we distinguish between strategic decisions, which relate to long-term goals and choices (Section~\ref{ssec:negotiating-collaboration-leadership}), and tactical decisions and actions needed to address any kind of imminent situation (Section~\ref{ssec:negotiating-collaboration-copilot}).

\subsubsection{Actions Separation}
\label{ssec:negotiating-collaboration-actions}
The division of actions between the pilot and the copilot can be static, without changes during a game, or dynamic.
It can be decided before starting, or during the game itself.
Furthermore, the actions can be separated strictly, with each player taking control of an exclusive subset of actions, or overlapping, with some actions controlled by both.

\hc{}
In general, \ParticipantOne{} notes that the division varies depending on the game. However, \ParticipantFour{} specifies that, when playing games belonging to the same genre, they often use similar configurations.
The division of actions is often defined before starting to play. For example, \ParticipantOne{} discusses the division with their copilot, since, having prior shared gaming experience, the copilot understands well what \ParticipantOne{} is better at.
Only \ParticipantTwo{} starts playing without a clear division of actions, preferring to work out together with the copilot what each of them does best as they play.
Usually, the action separation is static throughout the game.
\ParticipantTwo{}, instead, adopts a dynamic separation during the game: in specific sections that they would not be able to manage alone, the pilot hands over full control of the game to the copilot.
Most participants adopt a strict separation between the controls managed by the pilot and those managed by the copilot.
For example, \ParticipantOne{}, \ParticipantFour{}, and \ParticipantFive{} generally choose to delegate full camera control to the copilot.
\begin{displayquote}
\ParticipantFive{}: \it
One of \ParticipantFour{}'s biggest weaknesses actually is camera control on the right stick: using the camera while doing everything else is very, very hard. So, I'd say the thing I do most is camera control. [...] I'll point at an enemy and \ParticipantFour{} will go ahead and handle the job. 
\end{displayquote}
While strict separation of controls is more common, overlap between the pilot's and the copilot's controls is also possible.
This occurs mostly in family contexts or among children, where it becomes unclear who does what (\ParticipantSeven{}).
However, as \ParticipantFourteen{} points out, the lack of clarity on who is responsible for which controls can lead to confusion and frustration.

\pa{}
In partial automation, participants also expect the ability to customize which actions to automate and how they want to be supported.
As in human cooperation, \ParticipantFour{}, \ParticipantFive{}, \ParticipantSix{}, and \ParticipantSeven{} would prefer to divide the actions between pilot and copilot before starting to play.
\ParticipantFour{} assumes that the configuration would be similar to the one they already use when playing with a human copilot, delegating camera control to the software.
Instead, \ParticipantOne{} would experiment with different configurations to find the one that best suits their needs.
\ParticipantFour{} also hypothesizes the possibility of a dynamic division, as they would need the copilot's assistance only in specific sections of the game.
Finally, \ParticipantSix{} suggests asking the software copilot to propose how to divide the actions.
\ParticipantOne{} expresses a similar view but emphasizes that the copilot should not be overly intrusive in suggesting configuration changes or disrupting the flow of the game.
\begin{displayquote}
\ParticipantSix{}: \it
One of the things that is gonna be very important is that you understand the player, the person that you're building it for. [...] You might try to find some way to do an evaluation on the person [...], and then the AI could read what the evaluation says and say \quots{I think I already know what you need. You need the right trigger, the right bumper acceleration, and a boost}. 
\end{displayquote}


\subsubsection{Policies Guiding Action Assignment}
\label{ssec:negotiating-collaboration-policies}

\hc{}
We identified three main policies guiding the allocation of controls between pilot and copilot.
First, the pilot’s abilities are taken into account (\ParticipantOne{}, \ParticipantEight{}, \ParticipantNine{}, \ParticipantTen{}, and \ParticipantEleven{}), assigning to the copilot those controls that are less accessible to the pilot.
For example, \ParticipantOne{} has difficulties in controlling the left hand, and therefore remaps most frequently used game actions to buttons on the right side of the controller.
The remaining actions are then assigned to the copilot.
When both the pilot and the copilot have disabilities, the allocation is based on the abilities of both.
For instance, \ParticipantEight{}, \ParticipantNine{}, \ParticipantTen{}, and \ParticipantEleven{} all have a disability and regularly play in pairs using human cooperation, supporting each other during the game.
Similarly, \ParticipantSix{} reports having applied human cooperation multiple times with two people who both had disabilities.
%
Second, as hinted above by \ParticipantOne{}, the most important game actions are assigned to the pilot to ensure they maintain a sense of control over the game.
For example, \ParticipantFour{}, as the pilot, controls most commands, while \ParticipantFive{} assists with camera control, which \ParticipantFour{} cannot manage simultaneously with other actions.
Third, participants divide actions based on the macro-functionality they control.
For example, \ParticipantEight{} and \ParticipantNine{} separate movement from other actions, assigning all movement-related actions to one player and the remaining actions to the other.
%

\pa{}
The first two policies were also mentioned in relation to partial automation.
First, participants suggested delegating to the software copilot those actions that the pilot cannot control or would struggle to control.
For example, \ParticipantFour{} noted that partial automation could handle complex input sequences or those requiring quick execution. Similarly, actions requiring high precision, such as aiming (\ParticipantFour{}, \ParticipantTen{}), could also be delegated to the copilot. In this context, camera orientation was mentioned by many participants as a challenging action to manage and one that could be delegated to the copilot (\ParticipantOne{}, \ParticipantFour{}, \ParticipantEight{}).
Second, participants suggested delegating to the copilot secondary actions, in particular those used only occasionally (\ParticipantOne{}, \ParticipantFour{}), leaving the main controls to the pilot.


\subsubsection{Leadership Management}
\label{ssec:negotiating-collaboration-leadership}

\hc{}
In general, the person with disabilities acts as the pilot and makes game-wide strategic choices (\ParticipantTwo{}).
However, when both players have a disability, the definition of who is the pilot and who is the copilot becomes more blurred.
For example, \ParticipantEight{}, \ParticipantNine{}, \ParticipantTen{}, and \ParticipantEleven{} all have disabilities and play in pairs through human cooperation (Section~\ref{ssec:negotiating-collaboration-policies}).
Every decision is therefore made collaboratively, and both players contribute equally to the game.
Finally, \ParticipantSix{} notes that sometimes a person without disabilities can be the pilot, with a person with disabilities as the copilot.
\begin{displayquote}
\ParticipantSix{}: \it
Another patient that I had who also had hemiplegia [...] and he wanted to play Call of Duty~\cite{callOfDuty} with his son. And so [...] his son would look around and aim, and the father would just shoot.
\end{displayquote}
Leadership can also be temporarily assigned to the copilot.
\ParticipantTwo{} does this in game sections that require complex or rapid input sequences.
\ParticipantSix{} notes that when the copilot takes the lead, they must have experience in the specific game (Section~\ref{ssec:factors-affecting-collaboration-knowledge}), otherwise the experience could be frustrating for the pilot.

\pa{}
In partial automation, none of the participants mentioned the possibility of sharing or leaving game-wide strategic leadership to the software copilot.
Indeed, they believe that the pilot should maintain the leadership and partial automation should intervene only when the pilot is in difficulty, and generally avoid compromising their sense of control (\ParticipantOne{}, \ParticipantFour{}, \ParticipantSeven{}).
%
\ParticipantFour{} envisions leaving temporary control to the software copilot in specific contexts, as \ParticipantTwo{} does in human cooperation, for example, when the game requires action execution speed that is too high for them.
\begin{displayquote}
\ParticipantFour{}: \it
If the AI could change depending on the situation. I think that would be nice 'cause I hope I don't need the same level of help for an entire game. There are certain scenes that I need more help with than others.
\end{displayquote}


\subsubsection{Copilot's Operational Autonomy}
\label{ssec:negotiating-collaboration-copilot}

The copilot's level of operational autonomy exists on a spectrum.
At the one extreme, the copilot can act as an actuator, following step-by-step instructions from the pilot, who indicates when to act and which buttons to press.
At the other extreme, the copilot can be completely autonomous, executing actions based on a set of rules agreed upon with the pilot, without requiring specific instructions.

\hc{}
In human cooperation, \ParticipantOne{} is the only one who sometimes plays with a copilot acting solely as an actuator.
This is necessary when the copilot is not experienced with the game (Section~\ref{ssec:factors-affecting-collaboration-knowledge}), as reported by \ParticipantFourteen{}.
Instead, when the copilot knows the video game, they are able to act autonomously.
This results in the copilot providing more effective help, which causes the pilot to feel more immersed in the game (\ParticipantTwo{}).
As a result, most participants play with a copilot that acts autonomously (\ParticipantOne{}, \ParticipantFour{}, \ParticipantFive{}, \ParticipantSix{}, \ParticipantSeven{}, \ParticipantEight{}, \ParticipantNine{}, \ParticipantTen{}, and \ParticipantEleven{}).


\pa{}
In partial automation, \ParticipantTwo{}, \ParticipantFour{}, \ParticipantFive{}, \ParticipantTen{}, \ParticipantTwelve{}, and \ParticipantThirteen{} envision playing with a copilot that only follows their instructions.
\ParticipantTen{} specifies that they would use this support modality especially in logic games, where a more autonomous copilot might reveal the solution to proposed puzzles.
\ParticipantEight{} and \ParticipantNine{}, instead, would play with a copilot that autonomously controls a predefined set of actions.
\ParticipantOne{} and \ParticipantSix{} also anticipate that the copilot can act autonomously, but emphasize the importance of defining rules that precisely establish what it should and can do, for example, preventing the copilot from wasting ammunition in a first-person shooter game.

\subsection{Interaction}
\label{ssec:interaction}

The interaction between pilot and copilot in shared control is fundamental for exchanging information and coordinating during gameplay. 
This interaction is based both on communication, which can be verbal or non-verbal (Section~\ref{ssec:interaction-communication}), and on the ability to understand each other's intentions (Section~\ref{ssec:interaction-intent}).

\subsubsection{Verbal and Non-Verbal Communication}
\label{ssec:interaction-communication}

\hc{}
All participants interact primarily through verbal communication, engaging in dialogue with the other player to provide instructions and decide the strategy to adopt.
However, the way pilots and copilots communicate often evolves over time.
For example, by always collaborating with the same copilot, \ParticipantTwo{}, \ParticipantEight{}, and \ParticipantNine{} have developed short and quick commands over time to easily communicate with the copilot.
They all believe that this way of communicating is more efficient and less tiring than using full sentences.
Indeed, \ParticipantTwo{} recounts how their copilot acted as a guide using synthetic commands like \quots{clear left} or \quots{swoop} to indicate actions to perform in combat.
This way, \ParticipantTwo{} managed to complete some particularly complex game sequences requiring direct copilot intervention only in rare cases.
For \ParticipantTen{} and \ParticipantEleven{}, this type of communication is particularly important in faster-paced games for real-time coordination, while in slower games the communication is more articulate and focused on making decisions jointly.
Only \ParticipantEight{} also relies on non-verbal communication through eye contact, which they find more accessible due to speech difficulties.

\pa{}
Similarly, in partial automation, several participants propose using short and quick commands to give instructions to the copilot (\ParticipantEight{}, \ParticipantNine{}).
However, \ParticipantEight{} also highlights possible difficulties related to voice use, which could slow down interaction and, for those with speech problems, be less accessible.
For this reason, \ParticipantEight{} suggests teaching personalized commands to the software copilot.
Alternatively, \ParticipantOne{} suggests the use of brain-computer interfaces, which would enable immediate communication with the pilot, particularly suitable for fast-paced games.


\subsubsection{Intent Understanding}
\label{ssec:interaction-intent}

\hc{}
In human cooperation, the copilot's understanding of the pilot's intentions occurs primarily through explicit requests (\ParticipantSeven{}).
With experience, however, the copilot learns to anticipate the pilot's needs, becoming progressively more effective in providing support by observing the game and how the pilot plays (\ParticipantTwo{}).
Similarly, the pilot's understanding of the copilot's intentions is primarily based on verbal communication (\ParticipantTwo{}, \ParticipantFour{}, \ParticipantFive{}, \ParticipantSix{}).

\pa{}
In partial automation, mutual understanding of intentions remains equally important (\ParticipantSix{}). 
If the copilot fails to understand the pilot's objective, automatic intervention might be perceived as arbitrary. \ParticipantThirteen{} and \ParticipantFourteen{} agree on this issue, but also express doubts about the reliability of an automatic system in non-linear games, where possible actions are multiple and less predictable.
If, instead, the pilot fails to coordinate with the copilot, they might not be able to anticipate the copilot's actions and perceive a reduced sense of control over the game (\ParticipantSix{}).
\begin{displayquote}
\ParticipantSix{}:  \it
I think it would be tricky. If I had a shooting game and I wanted [the AI] to shoot for me, because I can't physically shoot the buttons. How do you control it from not using all of your ammunition? Will it just shoot all the time? Is there something that says \quots{Ok, now I have targeted my person, and now you shoot}?
\end{displayquote}
In this context, \ParticipantThirteen{} proposes a dynamic interaction model, where the copilot analyzes the pilot's choices and commands, proposing interactions consistent with their intentions, and improving over time the ability to anticipate user needs.



\subsection{Factors Affecting the Collaboration}
\label{ssec:factors-affecting-collaboration}

The collaboration between pilot and copilot is influenced by several factors that determine its effectiveness.
We first analyze how the copilot's knowledge of the game impacts the collaboration
(Section~\ref{ssec:factors-affecting-collaboration-knowledge}).
Then, we examine the impact of the relationship between the pilot and the copilot on in-game collaboration (Section~\ref{ssec:factors-affecting-collaboration-relation}).

\subsubsection{Knowledge of the Game}
\label{ssec:factors-affecting-collaboration-knowledge}


\hc{}
Participants emphasized that the copilot should have at least a general understanding of video games and their common mechanics.
\ParticipantSix{}, \ParticipantSeven{}, and \ParticipantTwelve{} consider this level of familiarity sufficient to provide effective support.
\ParticipantFive{} and \ParticipantSeven{} also mentioned specific mechanics that the copilot should be able to handle, such as camera control, character movement, and menu navigation.
For this reason, \ParticipantSix{} includes a training phase for the copilot when proposing human cooperation to players with disabilities.
In general, knowledge of the specific game for which support is provided is not deemed essential, but it can improve the gaming experience, particularly in complex video games (\ParticipantSix{}, \ParticipantFourteen{}, and \ParticipantFifteen{}).
For this reason, \ParticipantFive{} independently learns the game mechanics before assisting \ParticipantFour{} as copilot.
\ParticipantSix{} also considers experience important to facilitate communication during gameplay.
For \ParticipantOne{}, however, thorough knowledge of the game is essential when the copilot assumes a guiding role and needs to provide suggestions on how to proceed (Section~\ref{ssec:copilot-interventions-suggesting}).
In such cases, a lack of experience with the specific game may severely undermine the effectiveness of the assistance (\ParticipantSix{}), generating frustration in the pilot, who is unable to progress further.
\ParticipantTwo{} needs the copilot to describe what is happening on screen and therefore highlights that an experienced copilot better understands the state of the game and consequently communicates it more effectively.
\begin{displayquote}
\ParticipantTwo{}:
Let's say you get a person who knows a game well and one who doesn't know it as well. Like, one knows the enemy types and the other doesn't. [...] If my copilot's saying \quots{Okay, there's a terminator\footnote{\textit{Terminator} is a type of enemy in the video game \textit{Warhammer 40,000: Space Marine 2}} in the distance}, I'm like \quots{Okay, I know what to do} [...] Whereas if the person doesn't know the enemies as well you then have a sense of \quots{Oh God, this is terrifying. I don't know what's happening}.

\end{displayquote}

\pa{}
\ParticipantOne{} argues that a software copilot should be trained for each video game specifically to provide effective support.
However, \ParticipantOne{} also questions if developing a copilot capable of adapting to different games is feasible.


\subsubsection{Relationship Between Pilot and Copilot}
\label{ssec:factors-affecting-collaboration-relation}

\hc{} 
Participants reported that the copilot is often a family member or a friend.
For example, \ParticipantFour{} is a family member and caregiver of \ParticipantFive{}.
Due to this, \ParticipantFour{} is available to play together most of the time, and dedicates significant effort to better support \ParticipantFive{} (\textit{e.g.}, by trying a game in advance before playing together).
Furthermore, this type of relationship can facilitate communication, reduce logistical difficulties (Section~\ref{ssec:sc-limitations-availability}), and improve coordination during play.
For \ParticipantOne{}, \ParticipantEight{}, \ParticipantNine{}, \ParticipantTen{}, and \ParticipantEleven{} a friend acts as copilot.
\ParticipantOne{} highlights that prior gaming experience shared together facilitates collaboration.
For these reasons, experts (\ParticipantSix{}, \ParticipantSeven{}, \ParticipantFourteen{}, and \ParticipantFifteen{}), when recommending the use of human cooperation, suggest involving someone familiar to the pilot.
Familiarity is particularly important in fast-paced and complex video games (\ParticipantFour{}, \ParticipantEight{}, \ParticipantNine{}), whereas it is less critical in simpler games (\ParticipantTen{}, \ParticipantEleven{}).
Furthermore, \ParticipantSeven{} notes that some of the people they assist are reluctant to accept help from someone who doesn't already support them in daily life.
A copilot can also be a stranger met online.
This option was mentioned by \ParticipantTwo{}, who also observed that this type of copilot selection can lead to variable outcomes, depending on the compatibility with the copilot.
\begin{displayquote}
\ParticipantTwo{}: \it
There's this idea of being drift compatible\footnote{\textit{Drift compatible}: Term from the film Pacific Rim~\cite{pacificRim} in which two pilots jointly control a robot and need to be compatible to perform effectively} where you are in this flow state, you are both sort of synced up with each other, and it's all about shared communication.
\end{displayquote}
\ParticipantTwo{} further notes that each new collaboration requires time to build the coordination needed for a smooth and satisfying gameplay experience.
Indeed, the first hours of play with a new copilot are dedicated to deciding how to play together (Section~\ref{ssec:negotiating-collaboration}), communicating (Section~\ref{ssec:interaction}), and building a relationship.
During this process, errors and incomprehension are possible.
Thus, \ParticipantTwo{} and \ParticipantSix{} highlight that being positive and tolerant is important to avoid frustration.

\pa{}
As \ParticipantOne{} pointed out, the system should adapt its assistance to the player's skill level.
Otherwise, it can make the player dependent on the system, hindering autonomous growth and limiting their ability to experiment and autonomously find solutions.
Similarly, \ParticipantTwelve{}, and \ParticipantThirteen{} believe that the copilot should learn the pilot’s play style over time and adapt to it, as generic support may not be adequate or effective.



\section{Discussion}
\label{sec:discussion}

Prior works highlight that shared control can be used to access games that cannot be made accessible through existing accessibility solutions~\cite{cimolino2021role, gonccalves2023my}.
Our research confirms this (Section~\ref{ssec:sc-benefits})
and also uncovers \emph{which} gaming accessibility challenges are addressed through human cooperation, whether they can be addressed through partial automation as well, and how such support should be provided (Section~\ref{ssec:discussion-addressed_challenges}).
The resulting design recommendations are summarized in Table~\ref{tab:guidelines}.
%
In particular, we discuss advantages and disadvantages of partial automation with respect to human cooperation (Section~\ref{ssec:discussion-advantages}), focusing on the main limitation of human cooperation: the need for a human copilot.
This necessity limits the autonomy of players with disabilities, motivating partial automation as a substitute for human cooperation.
Then, based on how the pilot and the copilot collaborate in human cooperation, we discuss how this collaboration should translate into partial automation (Section~\ref{ssec:discussion-interaction}).
Finally, we analyze how the pilot and the copilot communicate during human cooperation, identifying possible feedback mechanisms and interactions that should be implemented in partial automation (Section~\ref{ssec:discussion-communication}).

\begin{table}[h]
\centering
\caption{Overview of partial automation design guidelines}
\label{tab:guidelines}
\begin{tabular}{l|l}
\hline
\multirow[c]{3}{*}{\bf Support in controlling game actions} 
& with fine control of input intensity \\
& with concurrent multi-dimensional controls \\
& with fast-paced control sequences \\ 
\hline
\multirow[c]{4}{*}{\bf Support by providing information} 
& allow on-demand requests for information \\
& detect when the player does not know how to progress \\
& provide hints without spoiling the fun of the player \\
& signal elements of interest in the game \\
\hline
\multirow[c]{3}{*}{\bf Support during game setup} 
& menu navigation support through voice commands \\
& facilities for explicit and personalized control assignment \\
& propose command split between pilot and copilot \\
\hline
\multirow[c]{2}{*}{\bf Support autonomy} 
& support for different platforms and games and be tuned on them \\
& reliable level of gaming experience at all times \\
\hline
\multirow[c]{2}{*}{\bf Support sociality} 
& also support human cooperation (possibly remotely) \\
& enable multiplayer assistance \\
\hline
\multirow[c]{2}{*}{\bf Subdivision of controls criteria} 
& agency maximization considering inputs the user can control \\
& clarity of actions separation: strict and by macro-functionality \\
\hline
\multirow[c]{4}{*}{\bf Copilot's operational autonomy} 
& allow on-demand requests by the player \\
& clearly defined autonomous interventions by the copilot \\
& learn when to assist, avoid doing too much and too well \\
& baseline multiplayer assistance / adaptive to player abilities \\
\hline
\multirow[c]{3}{*}{\bf Communication from the pilot} 
& natural language requests for explicit actions or suggestions \\
& personalized shorthand commands for fast-paced gaming \\
& non-verbal commands for people with speech impairment \\
\hline
\multirow[c]{2}{*}{\bf Feedback from the copilot} 
& verbal communication to signal/suggest something \\
& non-verbal feedback through screen, sounds, light, vibrations \\
\hline
\multirow[c]{2}{*}{\bf Mutual intent understanding} 
& pilot's intent: analysis of gameplay and external sensing \\
& copilot intent: feedback through visual, auditory, haptic cues \\
\hline
\end{tabular}
\Description{The table presents an overview of partial automation design guidelines. 
It is organized in two columns: the left column lists the main guideline categories, 
while the right column specifies the corresponding recommendations.}
\end{table}

\subsection{Gaming Accessibility Challenges Addressed by Shared Control}
\label{ssec:discussion-addressed_challenges}

The main challenge in controlling video games reported by our participants was when too many input dimensions needed to be controlled concurrently or in rapid succession.
Borrowing a term from the machine learning domain~\cite{keogh2011curse}, we call this issue the \emph{Curse of Dimensionality} (Section~\ref{ssub:discussion-addressed_challenges-curse}).
Unexpectedly, not all challenges addressed in shared control directly relate to control (\textit{i.e.,} pressing buttons on the controller).
Indeed, two additional challenges were also identified: the need for additional information on the game, what happens in it, and how to play it (Section~\ref{ssub:discussion-addressed_challenges-info}), and the difficulty in configuring the game setup before even playing the game (Section~\ref{ssub:discussion-addressed_challenges-setup}).

\subsubsection{Curse of Dimensionality}
\label{ssub:discussion-addressed_challenges-curse}

Prior works motivate the use of partial automation for game accessibility, arguing that: \quots{Games may require inputs that some players cannot provide with any device.}~\cite{cimolino2021role}.
Our analysis confirms this, since participants reported difficulties with \textbf{fine control of input intensity}, for example, when aiming in first-person shooters or steering in driving games (Section~\ref{ssec:negotiating-collaboration-policies}).
However, using existing accessibility tools and control remapping, most participants are able to mitigate this problem and can actually activate any single control individually.
Instead, our analysis uncovers the curse of dimensionality problem: the participants find it difficult to \textbf{control all actions concurrently} or in rapid succession.
Indeed, multiple participants reported difficulties when \textbf{simultaneously maneuvering multiple directional controls}, for example, the movement and the camera (Section~\ref{ssec:negotiating-collaboration-actions}).
The curse of dimensionality problem is further exacerbated when a large number of controls need to be managed during \textbf{fast-paced interaction sequences} (Section~\ref{ssec:negotiating-collaboration-leadership}).
Human cooperation is an effective solution to address the curse of dimensionality problem since it reduces the number of controls assigned to the pilot by delegating some to the copilot.
For the same reason, participants noted that partial automation could be similarly effective, highlighting its potential to replace human cooperation.


\subsubsection{Lack of Information}
\label{ssub:discussion-addressed_challenges-info}
In human cooperation, the copilot assists the pilot also by providing information throughout the game.
Partial automation can be used to replicate such support as well (Section~\ref{ssec:copilot-interventions-suggesting}).
Microsoft is currently investigating a similar type of support as part of their \textit{Copilot} conversational assistant~\cite{msGamingCopilot}, allowing players to explicitly ask suggestions about the game they are playing.
\textbf{Providing suggestions on-demand} is one way of providing information support, but partial automation should also be able to \textbf{detect when the user needs support} (\textit{e.g.}, does not know how to proceed) and \textbf{proactively provide suggestions}.
Such support can be useful to players with various needs, including individuals with cognitive impairments or beginners.
However, these \textbf{suggestions should be limited to what is needed} to progress, \textbf{without spoiling the game} for the pilot (Section~\ref{ssec:factors-affecting-collaboration-knowledge}).
A special type of suggestion, particularly relevant for accessibility, is \textbf{signaling important game elements} that the player may not have perceived (Section~\ref{ssec:copilot-interventions-signaling}).
This is highlighted by \ParticipantTwo{}, who is blind, but similar support could be used to signal important game sound cues to people with hearing impairments, as recently researched for real-world sound cues~\cite{huang2023not}.
This kind of support is reminiscent of Navi\footnote{Originally known as \quots{Fairy Navigation System} due to its purpose of providing guidance during the game.}, the fairy companion of the protagonist in Zelda, Ocarina of Time~\cite{zeldaOoT}, who alerts the player of dangers and elements of interest in the game.
%


\subsubsection{Support with the Game Setup}
\label{ssub:discussion-addressed_challenges-setup}
A human copilot also provides support during the game setup.
One type of support is the assistance in configuring the game setup, which can involve various hardware and software accommodations required to play a given game (Section~\ref{ssec:copilot-interventions-during-setup}).
This type of support may be needed even if the game itself can be played independently.
Clearly, partial automation cannot support the physical setup, which will still require human support.
Instead, it can provide support during software setup, particularly in deciding how to assign commands to the pilot and the copilot.
Indeed, while some participants expect the possibility to \textbf{decide control assignment explicitly}, others also suggested that partial automation could \textbf{propose how to subdivide the controls} (Section~\ref{ssec:negotiating-collaboration-actions}).
This is beneficial when the pilot has no prior knowledge of the game and, therefore, has no clue on how to divide the controls.
The software copilot can propose configurations based on the pilot's disability and previous gaming sessions (Section~\ref{ssec:negotiating-collaboration-actions}).

The second type of support is for navigating the game menu, which is a common accessibility problem reported by the participants (Section~\ref{ssec:copilot-interventions-menu-access}).
A partial automation system could \textbf{make the game menu more accessible}, acting as an actuator by \textbf{following the user's explicit instructions} (Section~\ref{ssec:negotiating-collaboration-copilot}).


\subsection{Advantages (and Disadvantages) of Partial Automation over Human Cooperation}
\label{ssec:discussion-advantages}

We highlight the positive impact of partial automation on the pilot's autonomy (Section~\ref{ssub:discussion-advantages-accessibility}), and we discuss the effects of human cooperation and partial automation on sociality, inclusivity, and engagement of the players (Section~\ref{ssub:discussion-advantages-sociality}).

\subsubsection{Autonomy}
\label{ssub:discussion-advantages-accessibility}

Participants highlighted one critical limitation of human cooperation: it requires the availability of a copilot.
An additional logistical constraint is that the pilot and the copilot need to be in the same place at the same time (Section~\ref{ssec:sc-limitations-availability}).
Remote human cooperation is one way to mitigate the constraint of being in the same place, but it still requires the availability of a human copilot and tools that are not widely known and are difficult to set up.
The copilot also needs to have adequate gaming abilities and knowledge of the game, both factors which influence the resulting gaming experience (Section~\ref{ssec:factors-affecting-collaboration-knowledge}).
Furthermore, at times, pilots felt that they might ask too much of copilots, and some completely gave up playing to avoid burdening the copilot (Section~\ref{ssec:sc-limitations-tradeoff}).
Another concern was the possible lack of engagement of the copilot in the game, since the agency is predominantly left to the pilot (Section~\ref{ssec:sc-limitations-engagement}).
While this specific concern seems unfounded, as the copilots we interviewed were eager to assist their pilots, it still discourages players with disabilities from asking for assistance.
All these barriers collectively impede players with disabilities who play in human cooperation from playing the games they want, when they want, and in the way they want (Section~\ref{ssec:sc-limitations-tradeoff}).

Due to these reasons, participants were enthusiastic about the prospect of a software copilot.
Of course, partial automation should be \textbf{available on different platforms and for different games} to be able to provide widespread and effective support.
Contrary to \ParticipantOne{}'s doubts on the feasibility of adapting partial automation to different games (Section~\ref{ssec:factors-affecting-collaboration-knowledge}), recent works~\cite{simateam2024scaling},
support the possibility of such a solution.
Once trained on a game, unlike human cooperation, partial automation would also \textbf{ensure a consistent level of support}, and therefore, of gaming experience.


\subsubsection{Sociality, Inclusivity, and Engagement}
\label{ssub:discussion-advantages-sociality}

Participants reported one limitation of the idea of substituting human cooperation with partial automation: the resulting loss of live social engagement (Section~\ref{ssec:sc-benefits-sociality}).
This is particularly true for those who organize social moments to play together and bond.
Human cooperation also gives the opportunity to play together with friends and family members.
Thus, participants considered partial automation not as a replacement for human cooperation, but as another option that would allow them more freedom to play in different ways.
So, participants foresee the \textbf{possibility of using either human cooperation or partial automation}, according to their preferences and the availability of a copilot.
Participants also note that partial automation would open up avenues for \textbf{multiplayer gaming} as another type of social interaction.


\subsection{Collaboration Between the Pilot and the Copilot}
\label{ssec:discussion-interaction}

Section~\ref{ssec:discussion-addressed_challenges} discusses what support can be provided by shared control.
Here, we discuss how such support should be provided.
Specifically, we discuss how the controls should be divided between the pilot and the copilot (Section~\ref{ssec:discussion-interaction-controls}) and the limits to when partial automation should intervene in the game (Section~\ref{ssec:discussion-interaction-interventions}).

\subsubsection{Subdivision of the Controls}
\label{ssec:discussion-interaction-controls}
A central aspect of shared control is determining how to divide the actions between the pilot and the copilot. We discuss the policies that should guide the configuration of a partial automation system and that could also be implemented by a partial automation system that automatically suggests a configuration. 

First, the subdivision should consider how many and \textbf{which inputs can the pilot control}.
Then, among all the actions in the game, the ones that provide the most sense of agency should be preferentially assigned to the input that the pilot can control, leaving the rest to the copilot.
In general, the actions that \textbf{maximize the sense of agency} are the ones that are more important for the gameplay or more frequently used (Section~\ref{ssec:negotiating-collaboration-policies}).
However, defining which actions are more important for the game is subjective; therefore, the process should allow for complete personalization.

Second, the subdivision of controls should take into account the \textbf{clarity of the subdivision}, avoiding sources of automation confusion~\cite{cimolino2023automation}.
Due to this, strict separation should be preferred (Section~\ref{ssec:negotiating-collaboration-actions}).
Indeed, if both the pilot and the copilot control a given action, it becomes unclear who is responsible for it, and it is harder for the player to understand the consequences of their own actions.
For the same reason, the actions belonging to the \textbf{same macro-functionality} should also have the same controller (Section~\ref{ssec:negotiating-collaboration-policies}).



\subsubsection{Tuning Copilot's Operational Autonomy}
\label{ssec:discussion-interaction-interventions}
Participants were concerned that partial automation interventions could compromise the pilot's sense of control.
Thus, they frequently expressed the need to limit the assistance solely to situations of necessity.
Interestingly, participants noted that human copilots that are not sufficiently autonomous can only act as actuators, thus spoiling the gaming experience (Section~\ref{ssec:negotiating-collaboration-copilot}).
In contrast, participants reported that if the software copilot is too skilled to play the game, it can reduce the pilot's sense of agency.
Thus, in some cases, they suggested limiting the software agent to responding only to the \textbf{pilot's explicit commands}.
This is suggested, for example, for puzzles or logic games, where the software agent solving the problem can spoil the game for the player (Section~\ref{ssec:negotiating-collaboration-copilot}).
Instead, for games characterized by fast-paced interactions, it could be unfeasible for the pilot to provide specific instructions, and thus the copilot should \textbf{intervene autonomously} (Section~\ref{ssub:discussion-addressed_challenges-curse}).
For example, if the software agent is responsible for controlling the orientation in a first-person shooter, it is impossible for the pilot to provide precise real-time commands (\textit{e.g.}, through voice) on how to adjust the orientation.
It should also be possible to combine explicit actions for which the copilot solely acts as an actuator with other actions in which the copilot is more autonomous.
For example, in a first-person shooter, the user can give explicit commands to change the weapon, while the software agent can autonomously aim at enemies.

In all cases, \textbf{action assignment must be clearly defined} in advance (Section~\ref{ssec:negotiating-collaboration-copilot}). 
Additionally, the software \textbf{copilot should not be too good}, leaving space for the player's sense of agency, and should \textbf{adapt to the player's changing abilities} (Section~\ref{ssec:copilot-interventions-playing}).
Autonomous interventions should also be \textbf{limited in multiplayer games} to avoid benefiting the user unfairly (Section~\ref{ssec:sc-limitations-ethical}).
To achieve this, the support should be set to a \textbf{baseline level of abilities} (Section~\ref{ssec:copilot-interventions-playing}).
Alternatively, the automated controls could be configured to \textbf{match the player's skills} (Section~\ref{ssec:sc-limitations-ethical}).


\subsection{Communication During the Gameplay}
\label{ssec:discussion-communication}
In human cooperation, the communication between the pilot and the copilot is fundamental.
Similarly, partial automation should also provide facilities to allow communication from the pilot (Section~\ref{ssec:discussion-communication-p2c}),
and feedback from the software copilot (Section~\ref{ssec:discussion-communication-c2p}).
Likewise, partial automation should also foster mutual intent understanding
(Section~\ref{ssec:discussion-communication-explicit-intent}).

\subsubsection{Communication From the Pilot}
\label{ssec:discussion-communication-p2c}
To support explicit requests from the pilot to the software copilot, partial automation needs to implement appropriate communication mechanisms.
The main use cases for such requests are for the pilot to request some actions to the copilot (Section~\ref{ssec:copilot-interventions-playing}), such as to change the weapon, or to ask the copilot for suggestions on how to proceed in the game (Section~\ref{ssec:copilot-interventions-suggesting}).
Participants were interested in making such requests \textbf{verbally using natural language} (Section~\ref{ssec:interaction-communication}).
Furthermore, they also suggested the possibility of teaching \textbf{personalized shorthand commands} to the system for a quicker and more comfortable interaction.

Some participants also highlighted the need to \textbf{communicate non-verbally}, for example, in the case of players with speech impairments.
For this purpose, various alternative communication channels could be used, such as head~\cite{manzoni2024personalized} or gaze~\cite{park2024functional} gestures, brain-computer interface~\cite{nama2024qc}, or non-verbal mouth sounds~\cite{ahmetovic2021replay}.
Although not explicitly mentioned by the participants, another possibility is to assign some of the game controller's buttons to the communication with the software agent. For example, we mentioned the possibility that the pilot verbally commands the agent to set the aim on the closest enemy on the right (Section~\ref{ssec:discussion-interaction-interventions}); an alternative solution is to use a button instead of a vocal command.
Note that this is not the same as directly controlling the aim, because in this case, the pilot only activates a single button, while the copilot handles the set of actions needed to target the enemy. 


\subsubsection{Feedback From the Copilot}
\label{ssec:discussion-communication-c2p}
The software copilot should also communicate with the pilot, possibly in \textbf{natural language}, to \textbf{signal elements of interest} in the game (Section~\ref{ssec:copilot-interventions-signaling}) and provide suggestions (Section~\ref{ssec:copilot-interventions-suggesting}).
The communication should take into account \textbf{user preferences}, such as the \textbf{user-defined shorthand terms}.
For effective natural language communication, Large Language Models~\cite{min2023recent} could be considered, with the prompt context taking into account information about the game and the user.
In addition to verbal communication, \textbf{non-verbal options} should be possible as well, for example, by displaying \textbf{text messages on the screen}.
\textbf{Symbolic communication}, through icons, \textbf{non-verbal sounds}, \textbf{light signals}, or \textbf{haptic feedback}, is also possible and should be configurable based on user preferences.
For example, the copilot could highlight elements of interest on the screen through a visual marker instead of verbally communicating the position of the element of interest.
Such awareness cues during gameplay are beneficial to the trust and reliance of the pilot towards the software copilot~\cite{cimolino2022impact}.


\subsubsection{Intent Understanding}
\label{ssec:discussion-communication-explicit-intent}
Aside from responding to explicit requests, partial automation should also attempt to \textbf{infer the user's intended actions} to provide more meaningful automated support.
Such an inference should be made by \textbf{observing the game} and the pilot's controls. Possibly, it could also consider \textbf{additional external cues}, like the pilot's \textbf{gaze and voice}.
However, understanding the user's intent can be challenging, in particular when the game does not have a clear goal (Section~\ref{ssec:interaction-intent}).
To address this issue, participants suggest that the copilot ask the pilot for confirmation on whether their intent is interpreted correctly. Of course, this should not interrupt the flow of the game. 

The system should also support the pilot to understand the copilot's actions, so that the pilot can also factor the copilot's actions into the decision-making process.
To this end, the copilot should provide \textbf{operational feedback} on what actions it is performing (Section~\ref{ssec:copilot-interventions-signaling}), possibly as \textbf{visual, haptic, or sound cues}.
This type of feedback can be considered a form of explainable AI~\cite{xu2019explainable}.


\subsection{Limitations of the Study}
In our study, we employed a convenience sampling method to recruit gamers and experts with experience in human cooperation for video game accessibility.
This approach allowed us to involve 14 representative participants.
However, we acknowledge that, due to the sampling approach used, there are some limitations to our research.
First, most of our participants have mobility impairments, with a notable exception of \ParticipantTwo{} who has a visual impairment.
Instead, we did not have participants with hearing or cognitive impairments.
Second, all our participants come from Western countries, which may introduce a cultural bias to our results.
Third, our study did not include participants who were very young or elderly.
These issues limit the generalizability of our findings, requiring future investigations to confirm our results.

Additionally, while all our participants had lived experience of human cooperation gaming, our findings are all based on their recounts.
A direct observational study would further help identify the factors that did not transpire solely from the interviews.
Furthermore, none of the participants had previously used a partial automation solution.
Thus, their opinions on partial automation are all hypothetical.
A possible extension of this work could therefore be a comparative observational study of human cooperation and partial automation, with the goal of confirming our current results.

\section{Conclusions and Future Work}
\label{sec:conclusions}
This work investigates current practices of shared control in video game accessibility through interviews and focus groups with players with disabilities, their supporters, and accessibility experts.
Findings indicate that shared control is a crucial means of enabling access to games that would otherwise remain unplayable, while also offering opportunities for social interaction and inclusion.
At the same time, reliance on human copilots introduces significant limitations, particularly regarding autonomy, availability, and engagement.
Participants expressed a strong interest in partial automation as a promising extension of shared control.
Such systems could increase independence, reduce logistical barriers, and broaden access.
Some concerns were raised as well, in particular about preserving player agency and maintaining the social value of cooperative play.
Accordingly, this study contributes to an understanding of how shared control is used in practice and how players perceive the transition toward automated copilots.
On this basis, we outline design guidelines that emphasize adaptability to individual needs and mechanisms for providing support while preserving the pilot's sense of agency.

Several directions for future research emerge from our findings.
First, the investigation could be extended to a broader and more diverse population of participants, including individuals with different types of disabilities, cultural backgrounds, and age groups. Such diversity would provide a more comprehensive picture of shared control practices and the potential generalizability of partial automation solutions.
Second, observational studies of real-world gaming sessions are needed to complement self-reported accounts and refine our understanding of how human cooperation unfolds in practice. Such studies could reveal additional design considerations and inform the development of automation features that more closely replicate or enhance human support.
Finally, the design guidelines identified in this work pave the way for prototyping and experimentally evaluating partial automation systems. Future studies could involve user testing of such prototypes, ideally in comparison with a human cooperation baseline, to assess their effectiveness, usability, and impact on player autonomy, agency, and enjoyment.

\bibliographystyle{ACM-Reference-Format}
\bibliography{main}


\begin{thebibliography}{64}


\ifx \showCODEN    \undefined \def \showCODEN     #1{\unskip}     \fi
\ifx \showDOI      \undefined \def \showDOI       #1{#1}\fi
\ifx \showISBNx    \undefined \def \showISBNx     #1{\unskip}     \fi
\ifx \showISBNxiii \undefined \def \showISBNxiii  #1{\unskip}     \fi
\ifx \showISSN     \undefined \def \showISSN      #1{\unskip}     \fi
\ifx \showLCCN     \undefined \def \showLCCN      #1{\unskip}     \fi
\ifx \shownote     \undefined \def \shownote      #1{#1}          \fi
\ifx \showarticletitle \undefined \def \showarticletitle #1{#1}   \fi
\ifx \showURL      \undefined \def \showURL       {\relax}        \fi
\providecommand\bibfield[2]{#2}
\providecommand\bibinfo[2]{#2}
\providecommand\natexlab[1]{#1}
\providecommand\showeprint[2][]{arXiv:#2}

\bibitem[8BitDo(nd)]%
        {liteSE}
\bibfield{author}{\bibinfo{person}{8BitDo}.} \bibinfo{year}{n.d.}\natexlab{}.
\newblock \bibinfo{booktitle}{\emph{8BitDo Lite SE}}.
\newblock
\urldef\tempurl%
\url{https://www.8bitdo.com/lite-se}
\showURL{%
\tempurl}


\bibitem[{Activision}(2003)]%
        {callOfDuty}
\bibfield{author}{\bibinfo{person}{{Activision}}.} \bibinfo{year}{2003}\natexlab{}.
\newblock \bibinfo{title}{Call of Duty}.
\newblock \bibinfo{howpublished}{Video game series}.
\newblock
\urldef\tempurl%
\url{https://www.callofduty.com}
\showURL{%
\tempurl}


\bibitem[Aguado-Delgado et~al\mbox{.}(2020)]%
        {aguado2020accessibility}
\bibfield{author}{\bibinfo{person}{Juan Aguado-Delgado}, \bibinfo{person}{Jos{\'e}-Maria Gutierrez-Martinez}, \bibinfo{person}{Jos{\'e}~R Hilera}, \bibinfo{person}{Luis De-Marcos}, {and} \bibinfo{person}{Salvador Ot{\'o}n}.} \bibinfo{year}{2020}\natexlab{}.
\newblock \showarticletitle{Accessibility in video games: a systematic review}.
\newblock \bibinfo{journal}{\emph{Universal Access in the Information Society}} \bibinfo{volume}{19}, \bibinfo{number}{1} (\bibinfo{year}{2020}), \bibinfo{pages}{169--193}.
\newblock


\bibitem[Ahmetovic et~al\mbox{.}(2021)]%
        {ahmetovic2021replay}
\bibfield{author}{\bibinfo{person}{Dragan Ahmetovic}, \bibinfo{person}{Daniele Riboli}, \bibinfo{person}{Cristian Bernareggi}, {and} \bibinfo{person}{Sergio Mascetti}.} \bibinfo{year}{2021}\natexlab{}.
\newblock \showarticletitle{RePlay: Touchscreen interaction substitution method for accessible gaming}. In \bibinfo{booktitle}{\emph{Proceedings of the 23rd International Conference on Mobile Human-Computer Interaction}}. \bibinfo{pages}{1--12}.
\newblock


\bibitem[Backes and Tso(1990)]%
        {backes1990umi}
\bibfield{author}{\bibinfo{person}{Paul~G Backes} {and} \bibinfo{person}{Kam~S Tso}.} \bibinfo{year}{1990}\natexlab{}.
\newblock \showarticletitle{UMI: An interactive supervisory and shared control system for telerobotics}. In \bibinfo{booktitle}{\emph{Proceedings., IEEE International Conference on Robotics and Automation}}. IEEE, \bibinfo{pages}{1096--1101}.
\newblock


\bibitem[Bierre et~al\mbox{.}(2005)]%
        {bierre2005game}
\bibfield{author}{\bibinfo{person}{Kevin Bierre}, \bibinfo{person}{Jonathan Chetwynd}, \bibinfo{person}{Barrie Ellis}, \bibinfo{person}{D~Michelle Hinn}, \bibinfo{person}{Stephanie Ludi}, {and} \bibinfo{person}{Thomas Westin}.} \bibinfo{year}{2005}\natexlab{}.
\newblock \showarticletitle{Game not over: Accessibility issues in video games}. In \bibinfo{booktitle}{\emph{Proc. of the 3rd International Conference on Universal Access in Human-Computer Interaction}}. \bibinfo{pages}{22--27}.
\newblock


\bibitem[Bierre et~al\mbox{.}(2004)]%
        {bierre2004accessibility}
\bibfield{author}{\bibinfo{person}{Kevin Bierre}, \bibinfo{person}{Michelle Hinn}, \bibinfo{person}{Teresa Martin}, \bibinfo{person}{Michael McIntosh}, \bibinfo{person}{Tess Snider}, \bibinfo{person}{Katie Stone}, {and} \bibinfo{person}{Thomas Westin}.} \bibinfo{year}{2004}\natexlab{}.
\newblock \bibinfo{booktitle}{\emph{Accessibility in Games: Motivations and Approaches}}.
\newblock \bibinfo{type}{White paper}. \bibinfo{institution}{International Game Developers Association Game Accessibility Special Interest Group}.
\newblock
\newblock
\shownote{White paper}.


\bibitem[Brown and Anderson(2021)]%
        {brown2021designing}
\bibfield{author}{\bibinfo{person}{Mark Brown} {and} \bibinfo{person}{Sky~LaRell Anderson}.} \bibinfo{year}{2021}\natexlab{}.
\newblock \showarticletitle{Designing for disability: Evaluating the state of accessibility design in video games}.
\newblock \bibinfo{journal}{\emph{Games and Culture}} \bibinfo{volume}{16}, \bibinfo{number}{6} (\bibinfo{year}{2021}), \bibinfo{pages}{702--718}.
\newblock


\bibitem[Brown and MacKenzie(2013)]%
        {brown2013evaluating}
\bibfield{author}{\bibinfo{person}{Michelle~A Brown} {and} \bibinfo{person}{I~Scott MacKenzie}.} \bibinfo{year}{2013}\natexlab{}.
\newblock \showarticletitle{Evaluating video game controller usability as related to user hand size}. In \bibinfo{booktitle}{\emph{Proceedings of the International Conference on Multimedia and Human Computer Interaction}}. \bibinfo{pages}{1--9}.
\newblock


\bibitem[ByoWave(nd)]%
        {proteusController}
\bibfield{author}{\bibinfo{person}{ByoWave}.} \bibinfo{year}{n.d.}\natexlab{}.
\newblock \bibinfo{booktitle}{\emph{Proteus Controller}}.
\newblock
\urldef\tempurl%
\url{https://byowave.com/products/proteus-controller}
\showURL{%
\tempurl}


\bibitem[Cechanowicz et~al\mbox{.}(2014)]%
        {cechanowicz2014improving}
\bibfield{author}{\bibinfo{person}{Jared~E Cechanowicz}, \bibinfo{person}{Carl Gutwin}, \bibinfo{person}{Scott Bateman}, \bibinfo{person}{Regan Mandryk}, {and} \bibinfo{person}{Ian Stavness}.} \bibinfo{year}{2014}\natexlab{}.
\newblock \showarticletitle{Improving player balancing in racing games}. In \bibinfo{booktitle}{\emph{Proceedings of the first ACM SIGCHI annual symposium on Computer-human interaction in play}}. \bibinfo{pages}{47--56}.
\newblock


\bibitem[Cimolino et~al\mbox{.}(2021)]%
        {cimolino2021role}
\bibfield{author}{\bibinfo{person}{Gabriele Cimolino}, \bibinfo{person}{Sussan Askari}, {and} \bibinfo{person}{TC~Nicholas Graham}.} \bibinfo{year}{2021}\natexlab{}.
\newblock \showarticletitle{The role of partial automation in increasing the accessibility of digital games}.
\newblock \bibinfo{journal}{\emph{Proceedings of the ACM on Human-Computer Interaction}} \bibinfo{volume}{5}, \bibinfo{number}{CHI PLAY} (\bibinfo{year}{2021}), \bibinfo{pages}{1--30}.
\newblock


\bibitem[Cimolino et~al\mbox{.}(2023)]%
        {cimolino2023automation}
\bibfield{author}{\bibinfo{person}{Gabriele Cimolino}, \bibinfo{person}{Renee Chen}, \bibinfo{person}{Carl Gutwin}, {and} \bibinfo{person}{TC~Nicholas Graham}.} \bibinfo{year}{2023}\natexlab{}.
\newblock \showarticletitle{Automation Confusion: A Grounded Theory of Non-Gamers’ Confusion in Partially Automated Action Games}. In \bibinfo{booktitle}{\emph{Proceedings of the 2023 CHI Conference on Human Factors in Computing Systems}}. \bibinfo{pages}{1--19}.
\newblock


\bibitem[Cimolino and Graham(2022)]%
        {cimolino2022two}
\bibfield{author}{\bibinfo{person}{Gabriele Cimolino} {and} \bibinfo{person}{TC~Nicholas Graham}.} \bibinfo{year}{2022}\natexlab{}.
\newblock \showarticletitle{Two heads are better than one: A dimension space for unifying human and artificial intelligence in shared control}. In \bibinfo{booktitle}{\emph{Proceedings of the 2022 CHI Conference on Human Factors in Computing Systems}}. \bibinfo{pages}{1--21}.
\newblock


\bibitem[Cimolino et~al\mbox{.}(2022)]%
        {cimolino2022impact}
\bibfield{author}{\bibinfo{person}{Gabriele Cimolino}, \bibinfo{person}{Carl Gutwin}, {and} \bibinfo{person}{T.C. Graham}.} \bibinfo{year}{2022}\natexlab{}.
\newblock \showarticletitle{Impact of Awareness Cues on Trust in Human-AI Shared Control}. In \bibinfo{booktitle}{\emph{TRAIT: Workshop on Trust and Reliance in AI–Human Teams, at CHI 2022}}. \bibinfo{publisher}{ACM}.
\newblock


\bibitem[ConsoleTuner(nd)]%
        {titanTwo}
\bibfield{author}{\bibinfo{person}{ConsoleTuner}.} \bibinfo{year}{n.d.}\natexlab{}.
\newblock \bibinfo{booktitle}{\emph{Titan Two}}.
\newblock
\urldef\tempurl%
\url{https://www.consoletuner.com/products/titan-two}
\showURL{%
\tempurl}


\bibitem[Controller(nd)]%
        {flexController}
\bibfield{author}{\bibinfo{person}{Flex Controller}.} \bibinfo{year}{n.d.}\natexlab{}.
\newblock \bibinfo{booktitle}{\emph{Flex Controller}}.
\newblock
\urldef\tempurl%
\url{https://www.flex-controller.com}
\showURL{%
\tempurl}


\bibitem[Dalgleish(2023)]%
        {dalgleish2023can}
\bibfield{author}{\bibinfo{person}{Mat Dalgleish}.} \bibinfo{year}{2023}\natexlab{}.
\newblock \showarticletitle{Who Can Play? Rethinking Video Game Controllers and Accessibility}.
\newblock In \bibinfo{booktitle}{\emph{Disability and Video Games: Practices of En-/Disabling Modes of Digital Gaming}}. \bibinfo{publisher}{Springer}, \bibinfo{pages}{43--71}.
\newblock


\bibitem[del Toro(2013)]%
        {pacificRim}
\bibfield{author}{\bibinfo{person}{Guillermo del Toro}.} \bibinfo{year}{2013}\natexlab{}.
\newblock \bibinfo{title}{Pacific Rim}.
\newblock \bibinfo{howpublished}{Warner Bros. Pictures and Legendary Pictures}.
\newblock
\newblock
\shownote{(Motion picture)}.


\bibitem[Dietz and Leigh(2001)]%
        {dietz2001diamondtouch}
\bibfield{author}{\bibinfo{person}{Paul Dietz} {and} \bibinfo{person}{Darren Leigh}.} \bibinfo{year}{2001}\natexlab{}.
\newblock \showarticletitle{DiamondTouch: a multi-user touch technology}. In \bibinfo{booktitle}{\emph{Proceedings of the 14th annual ACM symposium on User interface software and technology}}. \bibinfo{pages}{219--226}.
\newblock


\bibitem[Dragan and Srinivasa(2013)]%
        {dragan2013policy}
\bibfield{author}{\bibinfo{person}{Anca~D Dragan} {and} \bibinfo{person}{Siddhartha~S Srinivasa}.} \bibinfo{year}{2013}\natexlab{}.
\newblock \showarticletitle{A policy-blending formalism for shared control}.
\newblock \bibinfo{journal}{\emph{The International Journal of Robotics Research}} \bibinfo{volume}{32}, \bibinfo{number}{7} (\bibinfo{year}{2013}), \bibinfo{pages}{790--805}.
\newblock


\bibitem[EAD(2017)]%
        {marioKart}
\bibfield{author}{\bibinfo{person}{Nintendo EAD}.} \bibinfo{year}{2017}\natexlab{}.
\newblock \bibinfo{title}{Mario Kart 8 Deluxe}.
\newblock \bibinfo{howpublished}{Video game, released on Nintendo Switch}.
\newblock


\bibitem[Entertainment(2023)]%
        {ps5AccessController}
\bibfield{author}{\bibinfo{person}{Sony~Interactive Entertainment}.} \bibinfo{year}{2023}\natexlab{}.
\newblock \bibinfo{booktitle}{\emph{PS5 Access Controller}}.
\newblock
\urldef\tempurl%
\url{https://www.playstation.com/accessories/access-controller}
\showURL{%
\tempurl}


\bibitem[Feth et~al\mbox{.}(2009)]%
        {feth2009shared}
\bibfield{author}{\bibinfo{person}{Daniela Feth}, \bibinfo{person}{Binh~An Tran}, \bibinfo{person}{Raphaela Groten}, \bibinfo{person}{Angelika Peer}, {and} \bibinfo{person}{Martin Buss}.} \bibinfo{year}{2009}\natexlab{}.
\newblock \showarticletitle{Shared-control paradigms in multi-operator-single-robot teleoperation}.
\newblock In \bibinfo{booktitle}{\emph{Human Centered Robot Systems: Cognition, Interaction, Technology}}. \bibinfo{publisher}{Springer}, \bibinfo{pages}{53--62}.
\newblock


\bibitem[Gon{\c{c}}alves et~al\mbox{.}(2023)]%
        {gonccalves2023my}
\bibfield{author}{\bibinfo{person}{David Gon{\c{c}}alves}, \bibinfo{person}{Manuel Pi{\c{c}}arra}, \bibinfo{person}{Pedro Pais}, \bibinfo{person}{Jo{\~a}o Guerreiro}, {and} \bibinfo{person}{Andr{\'e} Rodrigues}.} \bibinfo{year}{2023}\natexlab{}.
\newblock \showarticletitle{" My Zelda Cane": Strategies Used by Blind Players to Play Visual-Centric Digital Games}. In \bibinfo{booktitle}{\emph{Proceedings of the 2023 CHI conference on human factors in computing systems}}. \bibinfo{pages}{1--15}.
\newblock


\bibitem[Gon{\c{c}}alves et~al\mbox{.}(2021)]%
        {gonccalves2021exploring}
\bibfield{author}{\bibinfo{person}{David Gon{\c{c}}alves}, \bibinfo{person}{Andr{\'e} Rodrigues}, \bibinfo{person}{Mike~L Richardson}, \bibinfo{person}{Alexandra~A De~Sousa}, \bibinfo{person}{Michael~J Proulx}, {and} \bibinfo{person}{Tiago Guerreiro}.} \bibinfo{year}{2021}\natexlab{}.
\newblock \showarticletitle{Exploring asymmetric roles in mixed-ability gaming}. In \bibinfo{booktitle}{\emph{Proceedings of the 2021 CHI Conference on Human Factors in Computing Systems}}. \bibinfo{pages}{1--14}.
\newblock


\bibitem[Gopinath et~al\mbox{.}(2016)]%
        {gopinath2016human}
\bibfield{author}{\bibinfo{person}{Deepak Gopinath}, \bibinfo{person}{Siddarth Jain}, {and} \bibinfo{person}{Brenna~D Argall}.} \bibinfo{year}{2016}\natexlab{}.
\newblock \showarticletitle{Human-in-the-loop optimization of shared autonomy in assistive robotics}.
\newblock \bibinfo{journal}{\emph{IEEE robotics and automation letters}} \bibinfo{volume}{2}, \bibinfo{number}{1} (\bibinfo{year}{2016}), \bibinfo{pages}{247--254}.
\newblock


\bibitem[Hougaard et~al\mbox{.}(2021)]%
        {hougaard2021willed}
\bibfield{author}{\bibinfo{person}{Bastian~Ils{\o} Hougaard}, \bibinfo{person}{Ingeborg~Goll Rossau}, \bibinfo{person}{Jedrzej~Jacek Czapla}, \bibinfo{person}{Mozes~Adorjan Miko}, \bibinfo{person}{Rasmus~Bugge Skammelsen}, \bibinfo{person}{Hendrik Knoche}, {and} \bibinfo{person}{Mads Jochumsen}.} \bibinfo{year}{2021}\natexlab{}.
\newblock \showarticletitle{Who willed it? decreasing frustration by manipulating perceived control through fabricated input for stroke rehabilitation BCI games}.
\newblock \bibinfo{journal}{\emph{Proceedings of the ACM on Human-Computer Interaction}} \bibinfo{volume}{5}, \bibinfo{number}{CHI PLAY} (\bibinfo{year}{2021}), \bibinfo{pages}{1--19}.
\newblock


\bibitem[Huang et~al\mbox{.}(2023)]%
        {huang2023not}
\bibfield{author}{\bibinfo{person}{Jeremy~Zhengqi Huang}, \bibinfo{person}{Hriday Chhabria}, {and} \bibinfo{person}{Dhruv Jain}.} \bibinfo{year}{2023}\natexlab{}.
\newblock \showarticletitle{“Not There Yet”: Feasibility and Challenges of Mobile Sound Recognition to Support Deaf and Hard-of-Hearing People}. In \bibinfo{booktitle}{\emph{Proceedings of the 25th International ACM SIGACCESS Conference on Computers and Accessibility}}. \bibinfo{pages}{1--14}.
\newblock


\bibitem[Hwang et~al\mbox{.}(2017)]%
        {hwang2017game}
\bibfield{author}{\bibinfo{person}{Susan Hwang}, \bibinfo{person}{Adrian L~Jessup Schneider}, \bibinfo{person}{Daniel Clarke}, \bibinfo{person}{Alexander Macintosh}, \bibinfo{person}{Lauren Switzer}, \bibinfo{person}{Darcy Fehlings}, {and} \bibinfo{person}{TC~Nicholas Graham}.} \bibinfo{year}{2017}\natexlab{}.
\newblock \showarticletitle{How game balancing affects play: Player adaptation in an exergame for children with cerebral palsy}. In \bibinfo{booktitle}{\emph{Proceedings of the 2017 conference on designing interactive systems}}. \bibinfo{pages}{699--710}.
\newblock


\bibitem[JoyToKey(nd)]%
        {joyToKey}
\bibfield{author}{\bibinfo{person}{JoyToKey}.} \bibinfo{year}{n.d.}\natexlab{}.
\newblock \bibinfo{booktitle}{\emph{JoyToKey}}.
\newblock
\urldef\tempurl%
\url{https://joytokey.net/en}
\showURL{%
\tempurl}


\bibitem[Keogh and Mueen(2011)]%
        {keogh2011curse}
\bibfield{author}{\bibinfo{person}{Eamonn Keogh} {and} \bibinfo{person}{Abdullah Mueen}.} \bibinfo{year}{2011}\natexlab{}.
\newblock \showarticletitle{Curse of dimensionality}.
\newblock In \bibinfo{booktitle}{\emph{Encyclopedia of machine learning}}. \bibinfo{publisher}{Springer}, \bibinfo{pages}{257--258}.
\newblock


\bibitem[Lazar et~al\mbox{.}(2017)]%
        {lazar2017research}
\bibfield{author}{\bibinfo{person}{Jonathan Lazar}, \bibinfo{person}{Jinjuan~Heidi Feng}, {and} \bibinfo{person}{Harry Hochheiser}.} \bibinfo{year}{2017}\natexlab{}.
\newblock \bibinfo{booktitle}{\emph{Research methods in human-computer interaction}}.
\newblock \bibinfo{publisher}{Morgan Kaufmann}.
\newblock


\bibitem[Li et~al\mbox{.}(2018)]%
        {li2018shared}
\bibfield{author}{\bibinfo{person}{Mingjun Li}, \bibinfo{person}{Haotian Cao}, \bibinfo{person}{Xiaolin Song}, \bibinfo{person}{Yanjun Huang}, \bibinfo{person}{Jianqiang Wang}, {and} \bibinfo{person}{Zhi Huang}.} \bibinfo{year}{2018}\natexlab{}.
\newblock \showarticletitle{Shared control driver assistance system based on driving intention and situation assessment}.
\newblock \bibinfo{journal}{\emph{IEEE Transactions on Industrial Informatics}} \bibinfo{volume}{14}, \bibinfo{number}{11} (\bibinfo{year}{2018}), \bibinfo{pages}{4982--4994}.
\newblock


\bibitem[Loparev et~al\mbox{.}(2014)]%
        {loparev2014introducing}
\bibfield{author}{\bibinfo{person}{Anna Loparev}, \bibinfo{person}{Walter~S. Lasecki}, \bibinfo{person}{Kyle~I. Murray}, {and} \bibinfo{person}{Jeffrey~P. Bigham}.} \bibinfo{year}{2014}\natexlab{}.
\newblock \showarticletitle{Introducing shared character control to existing video games}. In \bibinfo{booktitle}{\emph{International Conference on Foundations of Digital Games}}.
\newblock


\bibitem[Lu(2008)]%
        {lu2008evolution}
\bibfield{author}{\bibinfo{person}{William Lu}.} \bibinfo{year}{2008}\natexlab{}.
\newblock \showarticletitle{Evolution of video game controllers}.
\newblock \bibinfo{journal}{\emph{Roseville: Prima Publishing}}  \bibinfo{volume}{2} (\bibinfo{year}{2008}).
\newblock


\bibitem[Maggiorini et~al\mbox{.}(2017)]%
        {maggiorini2017evolution}
\bibfield{author}{\bibinfo{person}{Dario Maggiorini}, \bibinfo{person}{Marco Granato}, \bibinfo{person}{Laura~Anna Ripamonti}, \bibinfo{person}{Matteo Marras}, {and} \bibinfo{person}{Davide Gadia}.} \bibinfo{year}{2017}\natexlab{}.
\newblock \showarticletitle{Evolution of game controllers: Toward the support of gamers with physical disabilities}. In \bibinfo{booktitle}{\emph{International Conference on Computer-Human Interaction Research and Applications}}. Springer, \bibinfo{pages}{66--89}.
\newblock


\bibitem[Manzoni et~al\mbox{.}(2024)]%
        {manzoni2024personalized}
\bibfield{author}{\bibinfo{person}{Matteo Manzoni}, \bibinfo{person}{Dragan Ahmetovic}, {and} \bibinfo{person}{Sergio Mascetti}.} \bibinfo{year}{2024}\natexlab{}.
\newblock \showarticletitle{Personalized Facial Gesture Recognition for Accessible Mobile Gaming}. In \bibinfo{booktitle}{\emph{International Conference on Computers Helping People with Special Needs}}. Springer, \bibinfo{pages}{120--127}.
\newblock


\bibitem[Martinez et~al\mbox{.}(2024)]%
        {martinez2024playing}
\bibfield{author}{\bibinfo{person}{Jesse~J Martinez}, \bibinfo{person}{Jon~E Froehlich}, {and} \bibinfo{person}{James Fogarty}.} \bibinfo{year}{2024}\natexlab{}.
\newblock \showarticletitle{Playing on hard mode: Accessibility, difficulty and joy in video game adoption for gamers with disabilities}. In \bibinfo{booktitle}{\emph{Proceedings of the 2024 CHI Conference on Human Factors in Computing Systems}}. \bibinfo{pages}{1--17}.
\newblock


\bibitem[Medeiros and Coutinho(2015)]%
        {medeiros2015developing}
\bibfield{author}{\bibinfo{person}{Lucas Medeiros} {and} \bibinfo{person}{Flavio Coutinho}.} \bibinfo{year}{2015}\natexlab{}.
\newblock \showarticletitle{Developing an Accessible One-Switch Game for Motor Impaired Players}.
\newblock \bibinfo{journal}{\emph{Proceedings of SBGames}} (\bibinfo{year}{2015}), \bibinfo{pages}{236--239}.
\newblock


\bibitem[{Microsoft}(2017)]%
        {xboxshare}
\bibfield{author}{\bibinfo{person}{{Microsoft}}.} \bibinfo{year}{2017}\natexlab{}.
\newblock \bibinfo{booktitle}{\emph{Xbox Controller Assist (formerly Xbox Copilot)}}.
\newblock
\urldef\tempurl%
\url{https://support.xbox.com/en-US/help/account-profile/accessibility/copilot}
\showURL{%
\tempurl}


\bibitem[{Microsoft}(2025)]%
        {msGamingCopilot}
\bibfield{author}{\bibinfo{person}{{Microsoft}}.} \bibinfo{year}{2025}\natexlab{}.
\newblock \bibinfo{booktitle}{\emph{Gaming Copilot}}.
\newblock
\urldef\tempurl%
\url{https://news.xbox.com/2025/08/06/gaming-copilot-beta-begins-rolling-out-to-xbox-insiders-on-game-bar-today}
\showURL{%
\tempurl}


\bibitem[Microsoft(nd)]%
        {xboxAdaptiveController}
\bibfield{author}{\bibinfo{person}{Microsoft}.} \bibinfo{year}{n.d.}\natexlab{}.
\newblock \bibinfo{booktitle}{\emph{Xbox Adaptive Controller}}.
\newblock
\urldef\tempurl%
\url{https://www.xbox.com/accessories/controllers/xbox-adaptive-controller}
\showURL{%
\tempurl}


\bibitem[Min et~al\mbox{.}(2023)]%
        {min2023recent}
\bibfield{author}{\bibinfo{person}{Bonan Min}, \bibinfo{person}{Hayley Ross}, \bibinfo{person}{Elior Sulem}, \bibinfo{person}{Amir Pouran~Ben Veyseh}, \bibinfo{person}{Thien~Huu Nguyen}, \bibinfo{person}{Oscar Sainz}, \bibinfo{person}{Eneko Agirre}, \bibinfo{person}{Ilana Heintz}, {and} \bibinfo{person}{Dan Roth}.} \bibinfo{year}{2023}\natexlab{}.
\newblock \showarticletitle{Recent advances in natural language processing via large pre-trained language models: A survey}.
\newblock \bibinfo{journal}{\emph{Computing Surveys}} (\bibinfo{year}{2023}).
\newblock


\bibitem[Mosely et~al\mbox{.}(2022)]%
        {mosely2022video}
\bibfield{author}{\bibinfo{person}{Sarah Mosely}, \bibinfo{person}{Raeda Anderson}, \bibinfo{person}{George Usmanov}, \bibinfo{person}{John Morris}, {and} \bibinfo{person}{Ben Lippincott}.} \bibinfo{year}{2022}\natexlab{}.
\newblock \showarticletitle{Video game trends over time for people with disabilities}.
\newblock \bibinfo{journal}{\emph{The Journal on Technology and Persons with Disabilities}}  \bibinfo{volume}{232} (\bibinfo{year}{2022}).
\newblock


\bibitem[Nama and Samanta(2024)]%
        {nama2024qc}
\bibfield{author}{\bibinfo{person}{Tutan Nama} {and} \bibinfo{person}{Debasis Samanta}.} \bibinfo{year}{2024}\natexlab{}.
\newblock \showarticletitle{QC Speller: User Interface Design of a Hands-Free Touch-Free Speller with Brain Electroencephalogram Sensory Rhythm}.
\newblock \bibinfo{journal}{\emph{Transactions on Accessible Computing}} (\bibinfo{year}{2024}).
\newblock


\bibitem[{Nintendo}(1998)]%
        {zeldaOoT}
\bibfield{author}{\bibinfo{person}{{Nintendo}}.} \bibinfo{year}{1998}\natexlab{}.
\newblock \bibinfo{title}{The Legend of Zelda: Ocarina of Time}.
\newblock \bibinfo{howpublished}{Video game, originally released on Nintendo 64, and later on Nintendo GameCube}.
\newblock
\urldef\tempurl%
\url{https://www.nintendo.co.jp/n01/n64/software/zelda/index.html}
\showURL{%
\tempurl}


\bibitem[Park and Manduchi(2024)]%
        {park2024functional}
\bibfield{author}{\bibinfo{person}{Youn~Soo Park} {and} \bibinfo{person}{Roberto Manduchi}.} \bibinfo{year}{2024}\natexlab{}.
\newblock \showarticletitle{A Functional Usability Analysis of Appearance-Based Gaze Tracking for Accessibility}. In \bibinfo{booktitle}{\emph{Symposium on Eye Tracking Research and Applications}}.
\newblock


\bibitem[PlatinumGames(2014a)]%
        {bayonetta}
\bibfield{author}{\bibinfo{person}{PlatinumGames}.} \bibinfo{year}{2014}\natexlab{a}.
\newblock \bibinfo{title}{Bayonetta}.
\newblock \bibinfo{howpublished}{Video game, originally released on Xbox 360, PlayStation 3, and later on Nintendo Wii U, Nintendo Switch, Xbox One, PlayStation 4, and PC}.
\newblock


\bibitem[PlatinumGames(2014b)]%
        {bayonetta2}
\bibfield{author}{\bibinfo{person}{PlatinumGames}.} \bibinfo{year}{2014}\natexlab{b}.
\newblock \bibinfo{title}{Bayonetta 2}.
\newblock \bibinfo{howpublished}{Video game, originally released on Nintendo Wii U and later on Nintendo Switch}.
\newblock


\bibitem[PlayAbility(nd)]%
        {playAbility}
\bibfield{author}{\bibinfo{person}{PlayAbility}.} \bibinfo{year}{n.d.}\natexlab{}.
\newblock \bibinfo{booktitle}{\emph{PlayAbility}}.
\newblock
\urldef\tempurl%
\url{https://www.playability.gg}
\showURL{%
\tempurl}


\bibitem[{Priori Data}(2025)]%
        {gamersWorldwide}
\bibfield{author}{\bibinfo{person}{{Priori Data}}.} \bibinfo{year}{2025}\natexlab{}.
\newblock \bibinfo{booktitle}{\emph{How Many Gamers Are There in 2025? Latest Stats}}.
\newblock
\urldef\tempurl%
\url{https://prioridata.com/number-of-gamers/}
\showURL{%
\tempurl}


\bibitem[Rozendaal et~al\mbox{.}(2010)]%
        {rozendaal2010exploring}
\bibfield{author}{\bibinfo{person}{Marco~C Rozendaal}, \bibinfo{person}{Bram~AL Braat}, {and} \bibinfo{person}{Stephan~AG Wensveen}.} \bibinfo{year}{2010}\natexlab{}.
\newblock \showarticletitle{Exploring sociality and engagement in play through game-control distribution}.
\newblock \bibinfo{journal}{\emph{Ai \& Society}}  \bibinfo{volume}{25} (\bibinfo{year}{2010}), \bibinfo{pages}{193--201}.
\newblock


\bibitem[{SIMA Team} et~al\mbox{.}(2024)]%
        {simateam2024scaling}
\bibfield{author}{\bibinfo{person}{{SIMA Team}}, \bibinfo{person}{Maria~Abi Raad}, \bibinfo{person}{Arun Ahuja}, \bibinfo{person}{Catarina Barros}, \bibinfo{person}{Frederic Besse}, \bibinfo{person}{Andrew Bolt}, \bibinfo{person}{Adrian Bolton}, \bibinfo{person}{Bethanie Brownfield}, \bibinfo{person}{Gavin Buttimore}, \bibinfo{person}{Max Cant}, \bibinfo{person}{Sarah Chakera}, \bibinfo{person}{Stephanie C.~Y. Chan}, \bibinfo{person}{Jeff Clune}, \bibinfo{person}{Adrian Collister}, \bibinfo{person}{Vikki Copeman}, \bibinfo{person}{Alex Cullum}, \bibinfo{person}{Ishita Dasgupta}, \bibinfo{person}{Dario de Cesare}, \bibinfo{person}{Julia~Di Trapani}, \bibinfo{person}{Yani Donchev}, \bibinfo{person}{Emma Dunleavy}, \bibinfo{person}{Martin Engelcke}, \bibinfo{person}{Ryan Faulkner}, \bibinfo{person}{Frankie Garcia}, \bibinfo{person}{Charles Gbadamosi}, \bibinfo{person}{Zhitao Gong}, \bibinfo{person}{Lucy Gonzales}, \bibinfo{person}{Kshitij Gupta}, \bibinfo{person}{Karol Gregor}, \bibinfo{person}{Arne~Olav
  Hallingstad}, \bibinfo{person}{Tim Harley}, \bibinfo{person}{Sam Haves}, \bibinfo{person}{Felix Hill}, \bibinfo{person}{Ed Hirst}, \bibinfo{person}{Drew~A. Hudson}, \bibinfo{person}{Jony Hudson}, \bibinfo{person}{Steph Hughes-Fitt}, \bibinfo{person}{Danilo~J. Rezende}, \bibinfo{person}{Mimi Jasarevic}, \bibinfo{person}{Laura Kampis}, \bibinfo{person}{Rosemary Ke}, \bibinfo{person}{Thomas Keck}, \bibinfo{person}{Junkyung Kim}, \bibinfo{person}{Oscar Knagg}, \bibinfo{person}{Kavya Kopparapu}, \bibinfo{person}{Rory Lawton}, \bibinfo{person}{Andrew Lampinen}, \bibinfo{person}{Shane Legg}, \bibinfo{person}{Alexander Lerchner}, \bibinfo{person}{Marjorie Limont}, \bibinfo{person}{Yulan Liu}, \bibinfo{person}{Maria Loks-Thompson}, \bibinfo{person}{Joseph Marino}, \bibinfo{person}{Kathryn~Martin Cussons}, \bibinfo{person}{Loic Matthey}, \bibinfo{person}{Siobhan Mcloughlin}, \bibinfo{person}{Piermaria Mendolicchio}, \bibinfo{person}{Hamza Merzic}, \bibinfo{person}{Anna Mitenkova}, \bibinfo{person}{Alexandre
  Moufarek}, \bibinfo{person}{Valeria Oliveira}, \bibinfo{person}{Yanko Oliveira}, \bibinfo{person}{Hannah Openshaw}, \bibinfo{person}{Renke Pan}, \bibinfo{person}{Aneesh Pappu}, \bibinfo{person}{Alex Platonov}, \bibinfo{person}{Ollie Purkiss}, \bibinfo{person}{David Reichert}, \bibinfo{person}{John Reid}, \bibinfo{person}{Pierre~Harvey Richemond}, \bibinfo{person}{Tyson Roberts}, \bibinfo{person}{Giles Ruscoe}, \bibinfo{person}{Jaume~Sanchez Elias}, \bibinfo{person}{Tasha Sandars}, \bibinfo{person}{Daniel~P. Sawyer}, \bibinfo{person}{Tim Scholtes}, \bibinfo{person}{Guy Simmons}, \bibinfo{person}{Daniel Slater}, \bibinfo{person}{Hubert Soyer}, \bibinfo{person}{Heiko Strathmann}, \bibinfo{person}{Peter Stys}, \bibinfo{person}{Allison~C. Tam}, \bibinfo{person}{Denis Teplyashin}, \bibinfo{person}{Tayfun Terzi}, \bibinfo{person}{Davide Vercelli}, \bibinfo{person}{Bojan Vujatovic}, \bibinfo{person}{Marcus Wainwright}, \bibinfo{person}{Jane~X. Wang}, \bibinfo{person}{Zhengdong Wang}, \bibinfo{person}{Daan
  Wierstra}, \bibinfo{person}{Duncan Williams}, \bibinfo{person}{Nathaniel Wong}, \bibinfo{person}{Sarah York}, {and} \bibinfo{person}{Nick Young}.} \bibinfo{year}{2024}\natexlab{}.
\newblock \bibinfo{title}{Scaling Instructable Agents Across Many Simulated Worlds}.
\newblock
\newblock
\showeprint[arxiv]{2404.10179}~[cs.RO]
\urldef\tempurl%
\url{https://arxiv.org/abs/2404.10179}
\showURL{%
\tempurl}


\bibitem[Studios(2018)]%
        {kingdomComeDeliverance}
\bibfield{author}{\bibinfo{person}{Warhorse Studios}.} \bibinfo{year}{2018}\natexlab{}.
\newblock \bibinfo{title}{Kingdom Come: Deliverance}.
\newblock \bibinfo{howpublished}{Video game, released on Microsoft Windows, PlayStation 4, and Xbox One}.
\newblock


\bibitem[Sykownik et~al\mbox{.}(2017)]%
        {sykownik2017exploring}
\bibfield{author}{\bibinfo{person}{Philipp Sykownik}, \bibinfo{person}{Katharina Emmerich}, {and} \bibinfo{person}{Maic Masuch}.} \bibinfo{year}{2017}\natexlab{}.
\newblock \showarticletitle{Exploring patterns of shared control in digital multiplayer games}. In \bibinfo{booktitle}{\emph{International Conference on Advances in Computer Entertainment}}. Springer, \bibinfo{pages}{847--867}.
\newblock


\bibitem[Terry et~al\mbox{.}(2017)]%
        {terry2017thematic}
\bibfield{author}{\bibinfo{person}{Gareth Terry}, \bibinfo{person}{Nikki Hayfield}, \bibinfo{person}{Victoria Clarke}, \bibinfo{person}{Virginia Braun}, {et~al\mbox{.}}} \bibinfo{year}{2017}\natexlab{}.
\newblock \showarticletitle{Thematic analysis}.
\newblock \bibinfo{journal}{\emph{The SAGE handbook of qualitative research in psychology}} \bibinfo{volume}{2}, \bibinfo{number}{17-37} (\bibinfo{year}{2017}), \bibinfo{pages}{25}.
\newblock


\bibitem[{Turn 10 Studios}(2023)]%
        {forzaMotorsportAccessibility}
\bibfield{author}{\bibinfo{person}{{Turn 10 Studios}}.} \bibinfo{year}{2023}\natexlab{}.
\newblock \bibinfo{booktitle}{\emph{Forza Motorsport Accessibility Support}}.
\newblock
\urldef\tempurl%
\url{https://support.forzamotorsport.net/hc/en-us/articles/20964254277267-Forza-Motorsport-Accessibility-Support}
\showURL{%
\tempurl}


\bibitem[Vicencio-Moreira et~al\mbox{.}(2014)]%
        {vicencio2014effectiveness}
\bibfield{author}{\bibinfo{person}{Rodrigo Vicencio-Moreira}, \bibinfo{person}{Regan~L Mandryk}, \bibinfo{person}{Carl Gutwin}, {and} \bibinfo{person}{Scott Bateman}.} \bibinfo{year}{2014}\natexlab{}.
\newblock \showarticletitle{The effectiveness (or lack thereof) of aim-assist techniques in first-person shooter games}. In \bibinfo{booktitle}{\emph{Proceedings of the SIGCHI Conference on Human Factors in Computing Systems}}. \bibinfo{pages}{937--946}.
\newblock


\bibitem[VoiceAttack(nd)]%
        {voiceAttack}
\bibfield{author}{\bibinfo{person}{VoiceAttack}.} \bibinfo{year}{n.d.}\natexlab{}.
\newblock \bibinfo{booktitle}{\emph{VoiceAttack}}.
\newblock
\urldef\tempurl%
\url{https://voiceattack.com}
\showURL{%
\tempurl}


\bibitem[Wang et~al\mbox{.}(2020)]%
        {wang2020human}
\bibfield{author}{\bibinfo{person}{Dakuo Wang}, \bibinfo{person}{Elizabeth Churchill}, \bibinfo{person}{Pattie Maes}, \bibinfo{person}{Xiangmin Fan}, \bibinfo{person}{Ben Shneiderman}, \bibinfo{person}{Yuanchun Shi}, {and} \bibinfo{person}{Qianying Wang}.} \bibinfo{year}{2020}\natexlab{}.
\newblock \showarticletitle{From human-human collaboration to Human-AI collaboration: Designing AI systems that can work together with people}. In \bibinfo{booktitle}{\emph{Extended abstracts of the 2020 CHI conference on human factors in computing systems}}. \bibinfo{pages}{1--6}.
\newblock


\bibitem[Xbox(nd)]%
        {xboxAdaptiveJoystick}
\bibfield{author}{\bibinfo{person}{Xbox}.} \bibinfo{year}{n.d.}\natexlab{}.
\newblock \bibinfo{booktitle}{\emph{Xbox Adaptive Joystick}}.
\newblock
\urldef\tempurl%
\url{https://www.xbox.com/accessories/controllers/xbox-adaptive-joystick}
\showURL{%
\tempurl}


\bibitem[Xu et~al\mbox{.}(2019)]%
        {xu2019explainable}
\bibfield{author}{\bibinfo{person}{Feiyu Xu}, \bibinfo{person}{Hans Uszkoreit}, \bibinfo{person}{Yangzhou Du}, \bibinfo{person}{Wei Fan}, \bibinfo{person}{Dongyan Zhao}, {and} \bibinfo{person}{Jun Zhu}.} \bibinfo{year}{2019}\natexlab{}.
\newblock \showarticletitle{Explainable AI: A brief survey on history, research areas, approaches and challenges}. In \bibinfo{booktitle}{\emph{CCF international conference on natural language processing and Chinese computing}}. Springer, \bibinfo{pages}{563--574}.
\newblock


\bibitem[Yuan et~al\mbox{.}(2011)]%
        {yuan2011game}
\bibfield{author}{\bibinfo{person}{Bei Yuan}, \bibinfo{person}{Eelke Folmer}, {and} \bibinfo{person}{Frederick~C Harris~Jr}.} \bibinfo{year}{2011}\natexlab{}.
\newblock \showarticletitle{Game accessibility: a survey}.
\newblock \bibinfo{journal}{\emph{Universal Access in the information Society}} \bibinfo{volume}{10}, \bibinfo{number}{1} (\bibinfo{year}{2011}), \bibinfo{pages}{81--100}.
\newblock


\end{thebibliography}

\appendix

\end{document}